\documentclass[reprint,amsmath,amssymb,aps,prl]{revtex4-2}
\usepackage{graphicx}
\usepackage{dcolumn}% Align table columns on decimal point
\usepackage{bm}% bold math
\usepackage{blindtext}
\usepackage{xcolor}
\UseRawInputEncoding
\begin{document}

\preprint{APS/123-QED}

\title{Intrinsic Origin and Enhancement of Topological Responses in Ferrimagnetic Antiperovskite Mn$_4$N}
\author{Temuujin Bayaraa$^{1,2}$}
\email{tbayaraa@lbl.gov}
\author{Vsevolod Ivanov$^{1,3}$}
\author{Liang Z. Tan$^2$}
\author{Sin\'ead M. Griffin$^{1,2}$}
\affiliation{$^{1}$Materials Sciences Division, Lawrence Berkeley National Laboratory, Berkeley, California, 94720, USA}
\affiliation{$^{2}$Molecular Foundry Division, Lawrence Berkeley National Laboratory, Berkeley, California, 94720, USA}
\affiliation{$^{3}$Accelerator Technology and Applied Physics Division, Lawrence Berkeley National Laboratory, Berkeley, California, 94720, USA}

\date{\today}

\begin{abstract}

Using first-principles calculations we investigate the intrinsic origins of the anomalous Hall effect (AHE) and the anomalous Nernst effect (ANE) in antiperovskite ferrimagnet Mn$_4$N. We predict that the AHE is significantly enhanced under both compressive and tensile strain, however the ANE generally decreases under epitaxial strain, except for 1\% compressive strain. We connect this behavior to the evolution of the Berry curvature with strain, suggesting similar strategies for achieving large AHE and ANE changes with modest amounts of strain. Finally, we find that the non-monotonic characteristics of the AHE and ANE stem from the formation and movement of new Weyl points at the periphery of the Brillouin Zone under compressive and tensile strains.

\end{abstract}

\maketitle

The application of topology in condensed matter physics has become widely embraced and has renewed our understanding of electronic band structures of materials \cite{Hasan2010,Qi2011,Bansil2016}. This framework enables the understanding of symmetry-protected features in reciprocal space found in topological insulators and semimetals. Combining non-trivial topology with time-reversal symmetry breaking can lead to large Berry curvatures that enable sizable macroscopic responses such as the anomalous Hall effect (AHE) and the related anomalous Nernst effect (ANE) with great potential applications ranging from thermoelectrics to spin-based storage \cite{Manna2018,Xiao2010}. In fact, the key step to understanding the intrinsic origins of the AHE was in identifying the relationship between the AHE and the Berry curvature of the occupied electronic bands in a crystal \cite{Nagaosa2010}.

Antiperovskite transition-metal nitrides, especially Mn$_4$N, have a diverse range of magnetic properties and emergent phases which make them interesting for both understanding fundamental physics and for spin-based applications. Mn$_4$N has a high N\'{e}el temperature ($T_N$ = 745K), small saturation magnetization, and high uniaxial magnetic anisotropy, making it particularly appealing for thermoelectric applications based on the ANE \cite{Nakagawa1994,Ching1994,Yasutomi2014,Kabara2015,Isogami2020,Takei1960,Takei1962,Gushi2019,Ghosh2021,Isogami2021a,Isogami2021b,Isogami2022,Isogami2022a,Chen2022,Zhang2022a,Zhang2022b,Komori2019,Kabara2017a}. Mn$_4$N is also predicted to host a wealth of real-space magnetic topological features including spin textures, hedgehog-anti-hedgehog pairs and skyrmion tubes \cite{Ma2021, Wang2018, Isogami2022}. These non-trivial spin structures were found to be mainly stabilized by the frustration induced by the magnetic exchange interaction between fourth-nearest-neighbors \cite{Bayaraa2021a}. More recently, measurements of the AHE and ANE were reported for Mn$_4$N \cite{Kabara2017a,Shen2014,Isogami2021b,Isogami2022,Zhang2022a,Zhang2022b,Komori2019}, however, they do not agree on the origin of the AHE in Mn$_4$N, and importantly, do not address how it can be enhanced through experimentally viable routes such as strain. In particular, the microscopic origins of the AHE can either be extrinsic (e.g. due to spin-orbit induced scattering) or intrinsic (related to the Berry curvature). Recent experimental work studied transport signatures of the AHE in epitaxial Mn$_4$N films of different thickness, and concluded that the AHE has competing contributions from skew scattering, side jump, and intrinsic mechanisms \cite{Zhang2022a}. According to the conventional scaling law $\rho_{AHE}$ $\propto$ $\rho^{\gamma}_{xx}$ \cite{Nagaosa2010}, where $\rho_{AHE}$ is anomalous Hall resistivity and $\rho^{\gamma}_{xx}$ is longitudinal resistivity, $\gamma$ was found to be larger than 2 for all Mn$_4$N films, indicating that the side jump and intrinsic mechanisms are dominant in these films \cite{Karplus1954, Berger1970}. On the other hand, Isogami \textit{et al.} \cite{Isogami2021b} report a dominant intrinsic contribution to AHE and ANE based on transport and \textit{ab initio} calculations.

Surprisingly, we are not aware of a comprehensive study of the electronic origins of the AHE and ANE from the perspective of first-principles based calculations, or nor a discussion of how these properties can be enhanced. Moreover, the range of competing, frustrated magnetic states in ferrimagnetic Mn$_4$N motivates us to explore the range of tunability of the topological responses in this system. 

\begin{figure*}
\includegraphics[width=\textwidth]{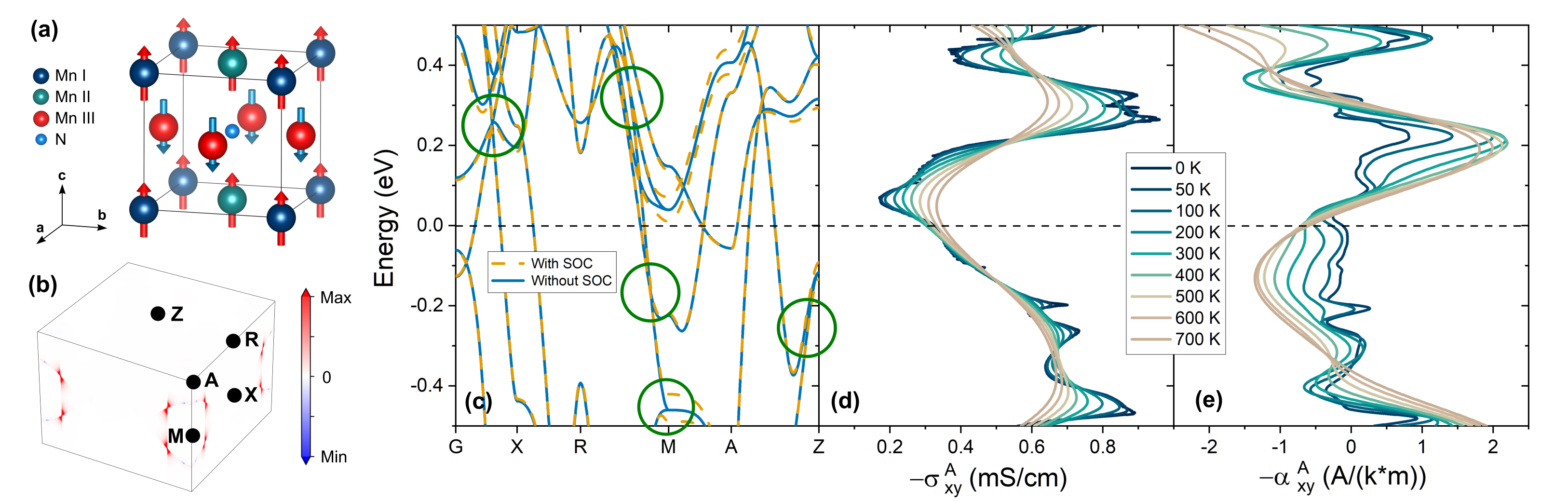}
\caption{The crystal structure of Mn$_4$N with Type-B magnetic structure  indicated with arrows (a), and the calculated Berry curvature, $\Omega_{z}$, distribution in the BZ (b). Black dots represent the high-symmetry points in the BZ and only most robust Berry curvature distributions are shown. Calculated band structure with and without spin–orbit coupling (c), calculated anomalous Hall conductivity, $\sigma_{xy}^A$ (d), and anomalous Nernst conductivity, $\alpha_{xy}^A$ (e), as a function of a total energy with respect to the Fermi energy. Note that green circles show the possible band-crossings that corresponds to the peaks in anomalous Hall effect. The Fermi level is marked by the dashed line.}
\label{fig1}
\end{figure*}

The antiperovskite structure of Mn$_4$N can be viewed as Mn$_3$MnN with Mn ions on three inequivalent cation sublattices and N taking the anion site (Fig.~\ref{fig1}(a)). These three different Mn sublattices have unequal magnetic moments leading to its ferrimagnetic nature and small saturation magnetization. Neutron diffraction experiments identified two different magnetic configurations in Mn$_4$N \cite{Ito2016} . In the ``Type-A" structure, the spins of Mn II and Mn III are aligned parallel to each other but antiparallel to those of Mn I, whereas in the ``Type-B" structure, the spins of Mn I and Mn II are aligned parallel to each other while being antiparallel to the spins of Mn III (see Fig.~\ref{fig1}(a)) \cite{Isogami2020,Isogami2021b,Bayaraa2021a}. Previous theoretical works found the Type-B to be the ground state, though both have been observed in experiment \cite{Isogami2021b}.
   
All first-principles calculations were carried out within the framework of Density Functional Theory (DFT) as implemented in Vienna \textit{Ab-Initio} Software Package (VASP) \cite{Kresse1999} using the projector augmented-wave potentials \cite{Blochl1994}. We used the generalized gradient approximation, with the Perdew-Burke-Ernzerhof exchange-correlation functional \cite{Liechtenstein1995,GGA}, and an effective Hubbard U parameter of 0.54 eV for the localized 3d electrons of the Mn ions. We selected this Hubbard U value so that the lattice parameters were close to those reported in experiment for the Type-B magnetic structure. In particular, we calculated the in-plane lattice constant, $a_{ip}$=3.897\AA~ ($a_{ip}$ = 3.89\AA~ in experiment \cite{Ito2016}) and the c/a axial ratio = 0.98 ($\approx $ 0.99 in experiment \cite{Ito2016}). All calculations were performed with the Type-B magnetic structure for which we found the calculated magnetic moments to be 3.6 $\mu_B$, 1.16 $\mu_B$, and -3.01 $\mu_B$ for Mn I, Mn II, and Mn III, respectively. We used an energy cutoff of 800 eV and a Monkhorst-Pack k-point mesh density of $13\times13\times13$ for Brillouin Zone (BZ) sampling. All structural relaxations were performed until the Hellmann-Feynman force on each atom is less than 0.001 eV/\AA. For the bulk case, we allowed all structural degrees of freedom to optimize (lattice parameters and internal coordinates), however for the epitaxial strain calculations, we fixed the in-plane (001) lattice vectors and allowed the out-of-plane lattice vector and atomic positions to relax. All calculations included spin-orbit coupling (SOC) self consistently as implemented in VASP. 

From the Bloch states obtained in the DFT calculation described above, we constructed a Wannier-based tight-binding model \cite{Wannier90} which we then used to calculate the intrinsic AHC ($\sigma_{ij}^{A}$) and ANE ($\alpha_{ij}^{A}$) as proposed by Xiao et al. \cite{ANE},

\begin{equation}
	\sigma_{ij}^{A} = - \frac{e^{2}}{\hbar} \sum_{n} \int \frac{dk}{(2\pi)^{3}} \Omega_{n,ij} f_{n}(T)
\end{equation}

\begin{eqnarray}
	\alpha_{ij}^{A} = - \frac{e}{T\hbar} \sum_{n} \int \frac{dk}{(2\pi)^{3}} \Omega_{n,ij} \lbrace (E_{n} - E_{F})f_{n}(T) + \nonumber \\
	k_{B}T \textrm{ln} (1 + e^{\frac{E_{n} - E_{F}}{-k_{B}T}}) \rbrace
\end{eqnarray}

where $e$ is the elementary charge, $\hbar$ the reduced Planck constant, $\Omega_{n,ij}$ the Berry curvature, $f_{n}$ the Fermi-Dirac distribution function with the band index \textit{n} and the wave vector \textit{\textbf{k}}, $k_{B}$ the Boltzmann constant, $E_{n}$ the band energy, and $E_{F}$ the Fermi level. We used a $501\times501\times501$ mesh for integrations over BZ to calculate the AHE and ANE in Wannier Tools \cite{Wanniertools}. Note that this mesh was carefully checked to ensure convergence.

We first focus on the calculated electronic topological properties of bulk Mn$_4$N. Fig.~\ref{fig1}(c) shows the calculated band structure with and without SOC between the high-symmetry points in the BZ. Note that the magnetic space group of Mn$_4$N is $P4/mm'm'$ with its magnetic easy axis along the out-of-plane z direction \cite{Stokes2005}. We find multiple band crossings without SOC that are gapped out with the inclusion of SOC, for instance along the R-M line. We identify these as the source of Berry curvature, $\Omega$, by explicit calculation of the Berry curvature where we plot its dominant component, $\Omega_Z$, in Fig.~\ref{fig1}(b). While there can be many different sources of $\Omega$, a strong contribution can originate from Weyl points (WP) that act as like a monopoles for $\Omega$, or from nodal lines that are gapped out when a mirror symmetry is broken via magnetization. We identify the WPs that contribute most to the Berry curvature using Wannier Tools \cite{Wanniertools}, and show them in the Supplemental Materials. 

Since such peaks in Berry curvature are an intrinsic source of the AHE (and ANE), we next report the calculated anomalous Hall conductivity (AHC) and anomalous Nernst conductivity (ANC) as functions of energy with respect to the Fermi level at temperatures ranging from 0 K to 700 K (Figs.~\ref{fig1}(d) and (e)) (Note that the N\'{e}el temperature of bulk Mn$_4$N is 745K \cite{Takei1960}). Since only the $\Omega_Z$ component of the Berry curvature is non-zero due to symmetry, we show the transverse (xy) component of the AHC ($\sigma_{xy}^A$) and the ANE ($\alpha_{xy}^A$), focusing on an energy range surrounding the Fermi level as this will determine the measured transport responses. We find that the band structure has several crossings above and below Fermi level that result in large Berry curvatures, leading to multiple peaks in the AHE shown in Fig.~\ref{fig1}(d). Here, we focus on the dominant peak at E $\sim$ +0.3 eV which corresponds to the band crossings shown in Fig.~\ref{fig1}(c). The AHE peak at E $\sim$ +0.3 eV becomes sharper as the system cools down and eventually splits into two peaks at E $\sim$ +0.28 eV and +0.32 eV at T = 0 K, corresponding to the band-crossings identified in the 0 K band structure in Fig.~\ref{fig1}(c).  On the other hand, we find that at high temperatures the ANE (Fig.~\ref{fig1}(e)) is peaked at E $\sim$ +0.2 eV and $\sim$ +0.4 eV but with opposite signs in this energy range close to the Fermi level. This is consistent with our calculated AHE peak at E $\sim$ +0.3 eV  since the thermoelectric conductivity, $\alpha_{ij}^A$, is proportional to the energy derivative of the AHE at low temperatures, which is known as the Mott relation. We note that the ANE signal at the Fermi level and the peak at E $\sim$ +0.2 eV both reach their maximum values around 400 K, dropping for higher temperatures. This is in contrast to the overall increase in thermoelectric conductivity with temperature, which reaches its maximum at 700 K for other energy ranges. This feature may be explained with the previous prediction of the magnetization compensation temperature ($\sim$500 K) in this system \cite{Bayaraa2021a}. However, we also note a sign reversal in the thermoelectric conductivity with increasing temperature at E $\sim$ +0.4 eV. While this is far enough away from the Fermi level in stoichiometric films to not influence transport, it could play a role in doped or alloyed systems \cite{Komori2020, Komori2022a, Komori2022b, Mitarai2020}.

We calculate the AHC, $\sigma_{xy}^A$, to be -323 S/cm at the Fermi level and at 0~K which is three times larger than the experimental result of -100 S/cm at 4 K \cite{Isogami2021b}. We note that our result is much closer to the experimental value than the previously reported calculated value of 573 S/cm \cite{Isogami2021b}, which may be due to different calculation methods. These include the choice of exchange-correlation functional, as well as the size of the k-mesh grid used to calculate AHE and ANE, as we found that particularly large mesh is required to reach a satisfactory convergence. The previous authors suggested the discrepancy between the calculated and measured $\sigma_{xy}^A$ was due to their samples being contaminated by a Type-A magnetic structure (which has negative $\sigma_{xy}^A$ according to their calculation results), resulting in an overall decrease. This same study also examined the possibility of an extrinsic mechanism accounting for the measured AHE response, leading to the conclusion that the observed AHE is mainly of intrinsic origin.

We next turn to the discussion of the ANE, where we find the calculated ANC, $\alpha_{xy}^A$, to be 0.65 Am$^{-1}$K$^{-1}$ at the Fermi level for 300 K where experiments report a value of 0.21 Am$^{-1}$K$^{-1}$ at 300 K \cite{Isogami2021b}. Similar to the calculated AHC, we find our calculated ANC value to be three times greater in magnitude than that measured in experiments. We find that the ANC takes its maximum of 0.7 Am$^{-1}$K$^{-1}$ at around 400 K which is comparable to other promising high-temperature ANE materials such as UCo$_{0.8}$Ru$_{0.2}$Al (15 Am$^{-1}$K$^{-1}$)\cite{Asaba2021}, Co$_3$Sn$_2$S$_2$ (10 Am$^{-1}$K$^{-1}$)\cite{Yang2020}, Co$_2$MnGa (4 Am$^{-1}$K$^{-1}$)\cite{Sakai2022}, Fe$_3$Sn (2.5 Am$^{-1}$K$^{-1}$)\cite{Chen2022}, SrRuO$_3$ (0.6 Am$^{-1}$K$^{-1}$)\cite{Miyasato2007}, and Mn$_3$Sn (0.3 Am$^{-1}$K$^{-1}$)\cite{Ikhlas2017}. Thus, Mn$_4$N can be a promising candidate for thermoelectric applications with its high N\'{e}el temperature, small saturation magnetization, and perpendicular magnetic anisotropy.

\begin{figure}

\includegraphics[width=0.5\textwidth]{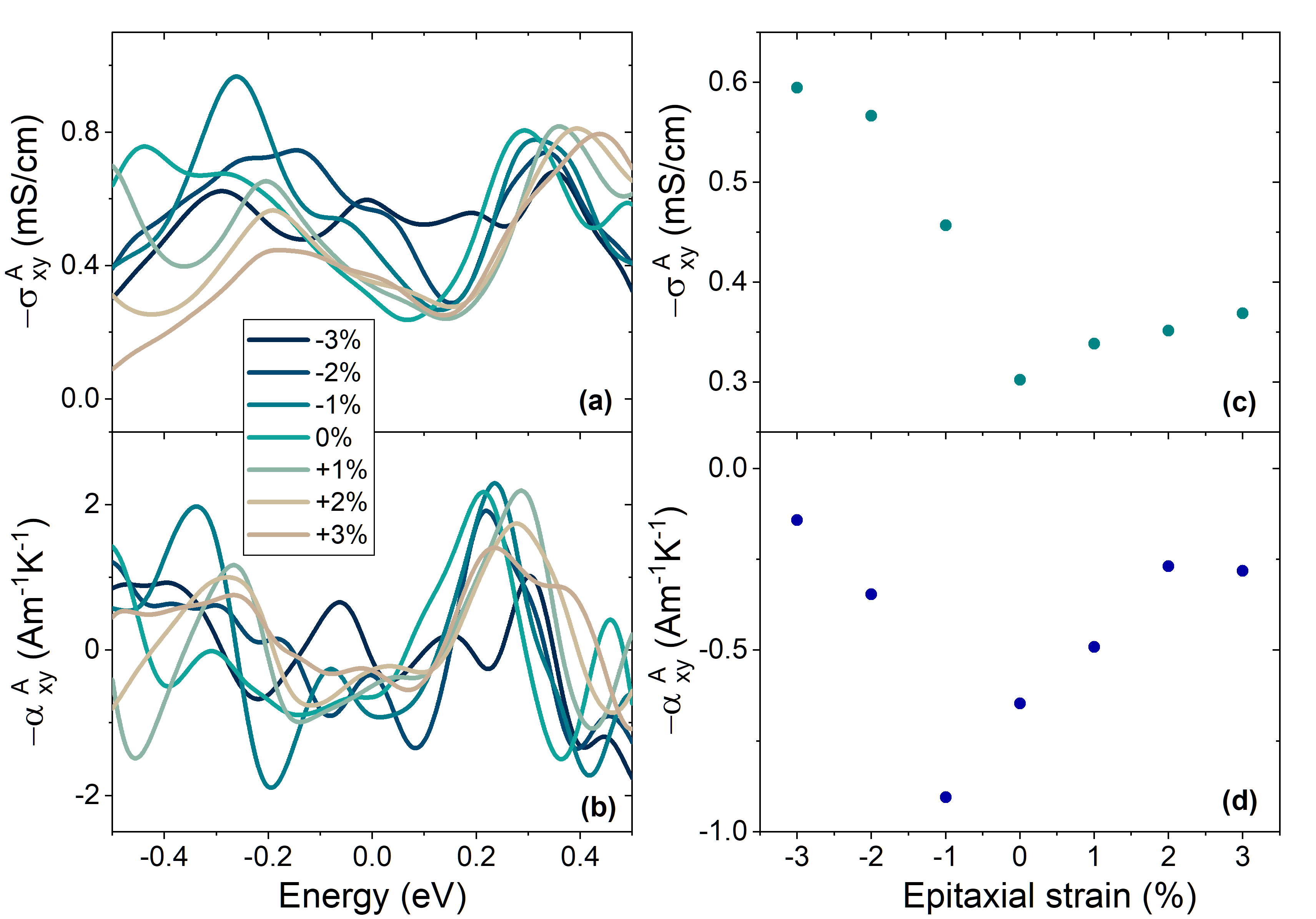}
\caption{Calculated aomalous Hall coefficient, $\sigma_{xy}^A$ (a) and anomalous Nernst coefficient, $\alpha_{xy}^A$ (b), as a function of a total energy with respect to the Fermi energy under different epitaxial strain at 300 K. Panels (c) and (d) shows the corresponding $\sigma_{xy}^A$ and $\alpha_{xy}^A$ at the Fermi level as a function of the applied epitaxial strain.}
\label{fig2}
\end{figure}

To explore routes to enhance the AHC and ANE, and to shed light on measurements of thin films, we calculate how epitaxial strain affects the AHE and ANE in Mn$_4$N \cite{Seva}. We first report the calculated $\sigma_{xy}^A$ and $\alpha_{xy}^A$ for both tensile and compressive strain in Fig.~\ref{fig2}(a) and (b) at 300 K, respectively. We find that the calculated values of the AHC and ANC at the Fermi level change significantly under epitaxial strain. We predict that both compressive and tensile strains increase the AHC almost linearly away from its turning point, with compressive strain having a much stronger effect (see Fig.~\ref{fig2}(c)).  For example, compressive strain increases $\sigma_{xy}^A$ from -302 S/cm (bulk) to -594 S/cm at -3\%, but tensile strain increases $\sigma_{xy}^A$ to only -369 S/cm at +3\%. However, we notice a change in the linear dependence of AHE with strain for -3\%, which shows a leveling off compared to smaller strains. Such results hint toward a change in Berry curvature distribution as the AHE is directly related to the Berry curvature as shown in Eq. (1), which we explore further below. In contrast to the enhancement of AHE with strain, we find the ANE decreases except for a compressive strain of 1\%, where we predict $\alpha_{xy}^A$ to be enhanced to a value of 0.9 Am$^{-1}$K$^{-1}$ from its bulk value of 0.65 Am$^{-1}$K$^{-1}$  (see Fig.~\ref{fig2}(d)). This is reasonable since $\alpha_{xy}^A$ is approximately proportional to $-d\sigma_{xy}^A/dE$ at low temperatures (see Eqs. 1 and 2).

\begin{figure}
\includegraphics[width=0.5\textwidth]{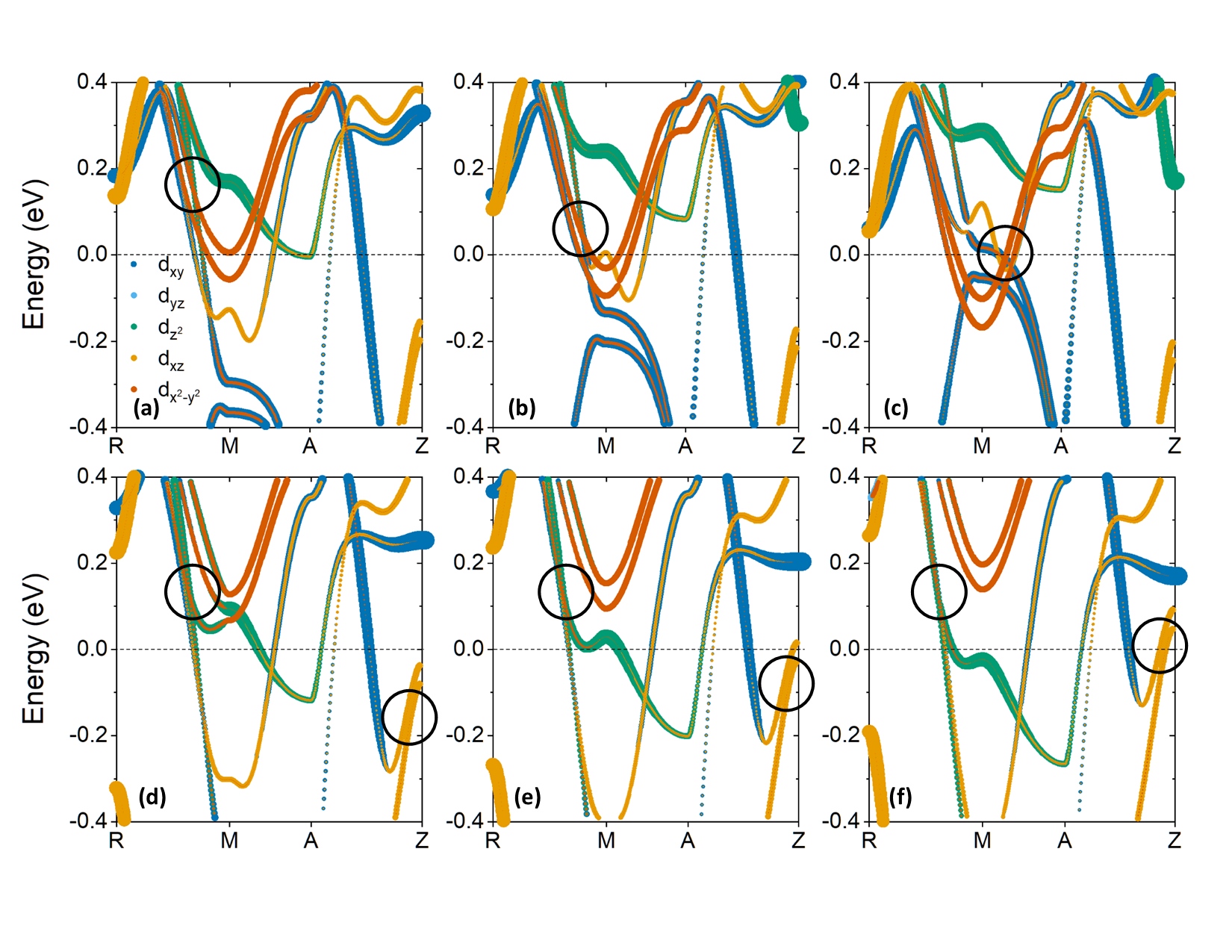}
\caption{Calculated band structures with Mn-\textit{d} orbitals projected as fat bands for epitaxial strains of -1\% (a), -2\% (b), -3\% (c), +1\% (d), +2\% (e), and +3\% (f), respectively. Spin-orbit coupling is included in all calculations with the Fermi level (dotted line) set to 0 eV. Black circles indicate the band-crossings that are the source of Berry curvature.} 
\label{fig3}
\end{figure}

\begin{figure}
\includegraphics[width=0.5\textwidth]{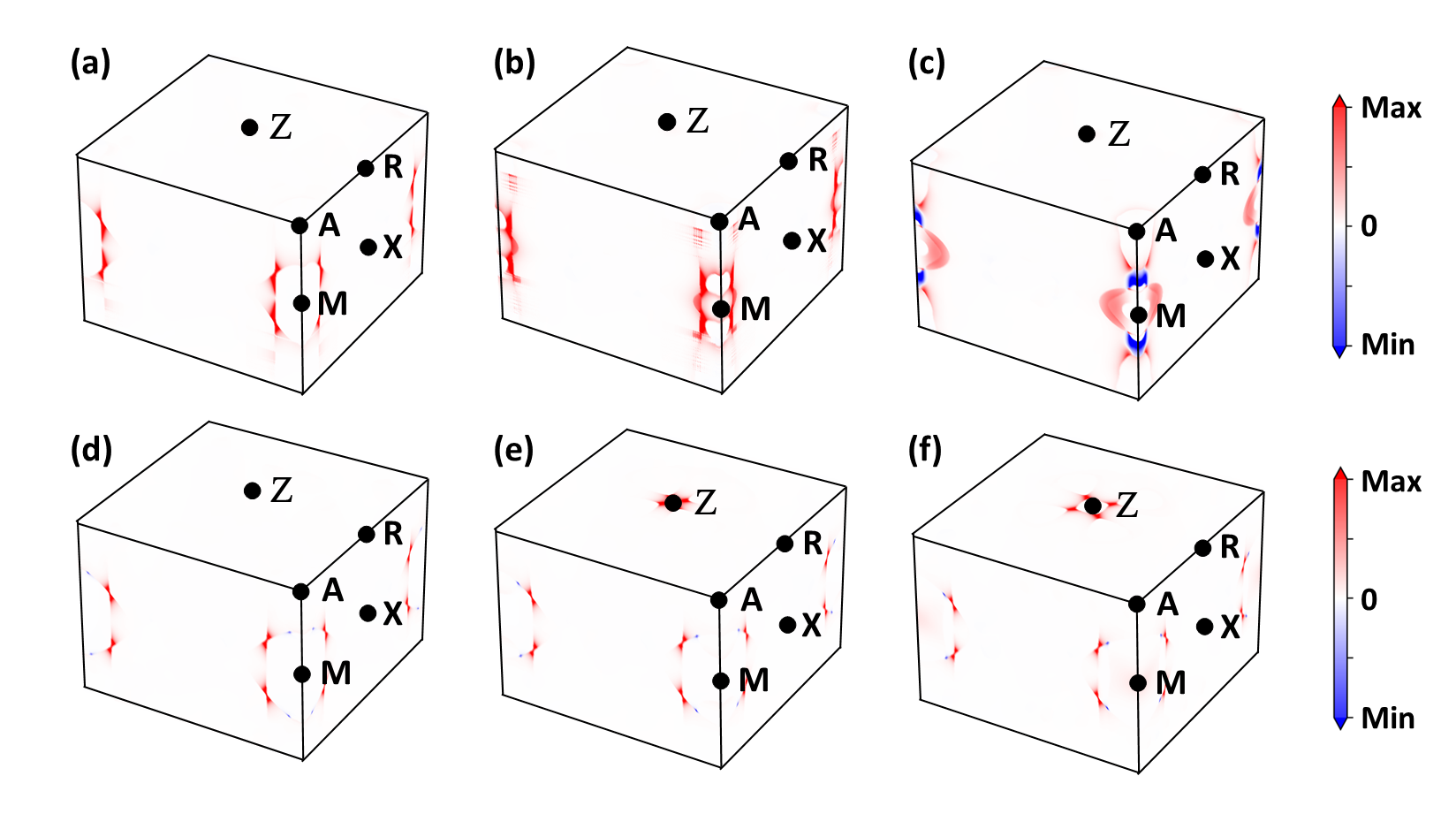}
\caption{Berry curvature, $\Omega_{z}$, distribution in the BZ for epitaxial strain values of -1\% (a), -2\% (b), -3\% (c), +1\% (d), +2\% (e), and +3\% (f), respectively. Note that only intense Berry curvature distributions are shown and black circles represent the high-symmetry k-points in the BZ.}
\label{fig4}
\end{figure}

To understand the intrinsic origins of the enhancement of AHE, and to explain the effect of epitaxial strain on the AHE and ANE, we next inspect our calculated electronic band structure for sources of Berry curvature and how they evolve with strain (Figs.~\ref{fig3} and \ref{fig4}).  On plotting the orbital-projected bands for the dominant Mn-\textit{d} orbitals in Fig.~\ref{fig3}, we find that the changes in AHE and ANE can be explained by the creation and movement of Weyl points near BZ edges.  Firstly, in the compressive strain region, we find that the band crossings that were responsible for the largest contribution to the AHE in the bulk case are systematically shifted closer to the Fermi level, from $\sim$+0.3 eV in the unstrained case (Fig.~\ref{fig1}) to $\sim$+0.15 eV (-1\% strain), $\sim$+0.05 eV (-2\% strain), and at the Fermi level (-3\% strain). This shift in the band crossings is due to both the shift of the Fermi level, and the change in orbital overlap with strain, primarily the relative downwards shift of the d$_{x^2-y^2}$ orbitals, and upwards shifts of the d$_{xy}$ and d$_{xz}$/d$_{yz}$ orbitals.
Closer inspection of the atom- and orbital-projections of these bands reveals that these overlapping bands are comprised mainly of the d$_{x^2-y^2}$, d$_{xy}$, d$_{xz}$ and d$_{yz}$ orbitals of Mn III (see Fig. S2 in the Supplemental Materials for atom-projected bands). The Mn III atoms form a square lattice in the Mn$_4$N structure: compressive strain greatly increases the orbital overlap of the in-plane orbitals (d$_{x^2-y^2}$ and d$_{xy}$), resulting in additional crossings in the M-A direction. We also find a slight increase in the overlap between those orbitals with an out-of-plane component (d$_{xz}$ and d$_{yz}$) due to the competition between the in-plane compression and out-of-plane expansion of those orbitals with compressive strain.  Illustrations depicting these changes are shown in Fig. S3 in the Supplemental Materials. We find the movement of our band crossings towards the Fermi level with strain to be consistent with our calculated Berry curvature, $\Omega_{z}$, which intensifies with increasing strain, as expected.  Crucially, we also find new regions of high $\Omega_{z}$ which appear at -3\% strain (Figs.~\ref{fig3}(c) and \ref{fig4}(c)), corresponding to the additional band crossings resulting from the increased overlap of the in-plane orbitals.  At -3\%, $\Omega_{z}$ has both positive and negative values which hints toward  the formation/creation of a new WP in addition to the WPs at lower compressive strains. The new WP is identified to originate from a new band-crossing along the M-A line (see Fig. S1) which is indicated in Fig.~\ref{fig3}(c) and a negative $\Omega_{z}$ distribution can be seen along the M-A line in Fig.~\ref{fig4}(c). Thus, with the formation of this new WP, the linear enhancement of AHE gets disrupted when going from -2\% to -3\% as seen in Fig.~\ref{fig2}(c).

On the other hand, in the tensile strain region, we also find band crossings shifting due to both the change in the Fermi level and orbital overlap under the tensile strain. In contrast to compressive strain, bands with d$_{x^2-y^2}$ orbital character shift upwards and bands with the d$_{z^2}$ and d$_{xz}$/d$_{yz}$ orbital character shift downwards. However, this downward shift of bands with d$_{xz}$/d$_{yz}$ orbital character is reversed near the high-symmetry point Z which can be understood from the increased overlap of d$_{xz}$ and d$_{yz}$ orbitals with tensile strain. As a result, another band-crossing that was below the Fermi level for the bulk case gets pushed up to the Fermi level, leading to the enhancement of AHE seen in Fig.~\ref{fig2}(c). This can also be seen from our calculated Berry curvature, $\Omega_{z}$, plots in Fig.~\ref{fig4}(d-f) where one can find new regions with high $\Omega_{z}$ between high-symmetry points A and Z that intensifies as more strain is applied.

In summary, we investigated the connection between the Berry curvature and the anomalous Hall and Nernst effects in the antiperovskite ferrimagnet Mn$_4$N with \textit{ab-initio}-based calculations for bulk and strained cases. A significant Berry curvature originating from WPs was found at the edge of the BZ. Epitaxial strain, both compressive and tensile, is predicted to enhance AHE, however, it is found to generally decrease ANE, except for the -1\%, for all studied strains. At a compressive strain of -3\%, new WPs are predicted to form originating from new band crossings induced by orbital overlap. Overall, our results show that enhancement of AHE is due to two factors (i) shifting of the relative Fermi level closer to band crossings, and (ii) the creation of new band crossing with changes of orbital overlap under strain. When taken in conjunction with previous works on real-space topology \cite{Bayaraa2021a}, such predictions thus indicate that Mn$_4$N is topological in both real and reciprocal spaces, and could provide an interesting system to explore the interplay of robust topological features in such doubly-topological magnetic systems \cite{Weber2019}. Furthermore, a large enhancement of AHE with compressive strain may improve the stability of the metastable phases in real-space since the topological Hall effect relation to AHE is reported to increase as temperature lowers \cite{Isogami2022} and topological meta-stable phases were predicted at low temperatures \cite{Bayaraa2021a}. We hope our prediction will motivate experimental studies on Mn$_4$N thin films and confirm our findings of epitaxial strain effect on AHE and ANE, and will be put to use to design novel devices.

This work is supported by the U.S. Department of Energy, Office of Science, Office of Basic Energy Sciences, Materials Sciences and Engineering Division under Contract No. DE-AC02-05-CH11231 within the Nonequilibrium Magnetic Materials Program (MSMAG). Computational resources were provided by the National Energy Research Scientific Computing Center and the Molecular Foundry, DOE Office of Science User Facilities supported by the Office of Science, U.S. Department of Energy under Contract No. DEAC02-05CH11231. The work performed at the Molecular Foundry was supported by the Office of Science, Office of Basic Energy Sciences, of the U.S. Department of Energy under the same contract.

% \newpage

\bibliographystyle{apsrev}
\bibliography{library}

\begin{thebibliography}{50}
\expandafter\ifx\csname natexlab\endcsname\relax\def\natexlab#1{#1}\fi
\expandafter\ifx\csname bibnamefont\endcsname\relax
  \def\bibnamefont#1{#1}\fi
\expandafter\ifx\csname bibfnamefont\endcsname\relax
  \def\bibfnamefont#1{#1}\fi
\expandafter\ifx\csname citenamefont\endcsname\relax
  \def\citenamefont#1{#1}\fi
\expandafter\ifx\csname url\endcsname\relax
  \def\url#1{\texttt{#1}}\fi
\expandafter\ifx\csname urlprefix\endcsname\relax\def\urlprefix{URL }\fi
\providecommand{\bibinfo}[2]{#2}
\providecommand{\eprint}[2][]{\url{#2}}

\bibitem[{\citenamefont{Hasan and Kane}(2010)}]{Hasan2010}
\bibinfo{author}{\bibfnamefont{M.~Z.} \bibnamefont{Hasan}} \bibnamefont{and}
  \bibinfo{author}{\bibfnamefont{C.~L.} \bibnamefont{Kane}},
  \bibinfo{journal}{Reviews of Modern Physics} \textbf{\bibinfo{volume}{82}},
  \bibinfo{pages}{3045} (\bibinfo{year}{2010}), ISSN \bibinfo{issn}{00346861},
  \urlprefix\url{https://journals.aps.org/rmp/abstract/10.1103/RevModPhys.82.3045}.

\bibitem[{\citenamefont{Qi and Zhang}(2011)}]{Qi2011}
\bibinfo{author}{\bibfnamefont{X.~L.} \bibnamefont{Qi}} \bibnamefont{and}
  \bibinfo{author}{\bibfnamefont{S.~C.} \bibnamefont{Zhang}},
  \bibinfo{journal}{Reviews of Modern Physics} \textbf{\bibinfo{volume}{83}},
  \bibinfo{pages}{1057} (\bibinfo{year}{2011}), ISSN \bibinfo{issn}{00346861},
  \eprint{1008.2026},
  \urlprefix\url{https://journals.aps.org/rmp/abstract/10.1103/RevModPhys.83.1057}.

\bibitem[{\citenamefont{Bansil et~al.}(2016)\citenamefont{Bansil, Lin, and
  Das}}]{Bansil2016}
\bibinfo{author}{\bibfnamefont{A.}~\bibnamefont{Bansil}},
  \bibinfo{author}{\bibfnamefont{H.}~\bibnamefont{Lin}}, \bibnamefont{and}
  \bibinfo{author}{\bibfnamefont{T.}~\bibnamefont{Das}},
  \bibinfo{journal}{Reviews of Modern Physics} \textbf{\bibinfo{volume}{88}},
  \bibinfo{pages}{021004} (\bibinfo{year}{2016}), ISSN
  \bibinfo{issn}{15390756}, \eprint{1603.03576},
  \urlprefix\url{https://journals.aps.org/rmp/abstract/10.1103/RevModPhys.88.021004}.

\bibitem[{\citenamefont{Manna et~al.}(2018)\citenamefont{Manna, Sun, Muechler,
  K{\"{u}}bler, and Felser}}]{Manna2018}
\bibinfo{author}{\bibfnamefont{K.}~\bibnamefont{Manna}},
  \bibinfo{author}{\bibfnamefont{Y.}~\bibnamefont{Sun}},
  \bibinfo{author}{\bibfnamefont{L.}~\bibnamefont{Muechler}},
  \bibinfo{author}{\bibfnamefont{J.}~\bibnamefont{K{\"{u}}bler}},
  \bibnamefont{and} \bibinfo{author}{\bibfnamefont{C.}~\bibnamefont{Felser}},
  \bibinfo{journal}{Nature Reviews Materials 2018 3:8}
  \textbf{\bibinfo{volume}{3}}, \bibinfo{pages}{244} (\bibinfo{year}{2018}),
  ISSN \bibinfo{issn}{2058-8437}, \eprint{1802.03771},
  \urlprefix\url{https://www.nature.com/articles/s41578-018-0036-5}.

\bibitem[{\citenamefont{Xiao et~al.}(2010)\citenamefont{Xiao, Chang, and
  Niu}}]{Xiao2010}
\bibinfo{author}{\bibfnamefont{D.}~\bibnamefont{Xiao}},
  \bibinfo{author}{\bibfnamefont{M.~C.} \bibnamefont{Chang}}, \bibnamefont{and}
  \bibinfo{author}{\bibfnamefont{Q.}~\bibnamefont{Niu}},
  \bibinfo{journal}{Reviews of Modern Physics} \textbf{\bibinfo{volume}{82}},
  \bibinfo{pages}{1959} (\bibinfo{year}{2010}), ISSN \bibinfo{issn}{00346861},
  \eprint{0907.2021},
  \urlprefix\url{https://journals.aps.org/rmp/abstract/10.1103/RevModPhys.82.1959}.

\bibitem[{\citenamefont{Nagaosa et~al.}(2010)\citenamefont{Nagaosa, Sinova,
  Onoda, MacDonald, and Ong}}]{Nagaosa2010}
\bibinfo{author}{\bibfnamefont{N.}~\bibnamefont{Nagaosa}},
  \bibinfo{author}{\bibfnamefont{J.}~\bibnamefont{Sinova}},
  \bibinfo{author}{\bibfnamefont{S.}~\bibnamefont{Onoda}},
  \bibinfo{author}{\bibfnamefont{A.~H.} \bibnamefont{MacDonald}},
  \bibnamefont{and} \bibinfo{author}{\bibfnamefont{N.~P.} \bibnamefont{Ong}},
  \bibinfo{journal}{Reviews of Modern Physics} \textbf{\bibinfo{volume}{82}},
  \bibinfo{pages}{1539} (\bibinfo{year}{2010}), ISSN \bibinfo{issn}{00346861},
  \eprint{0904.4154}.

\bibitem[{\citenamefont{Nakagawa and Naoe}(1994)}]{Nakagawa1994}
\bibinfo{author}{\bibfnamefont{S.}~\bibnamefont{Nakagawa}} \bibnamefont{and}
  \bibinfo{author}{\bibfnamefont{M.}~\bibnamefont{Naoe}},
  \bibinfo{journal}{Journal of Applied Physics} \textbf{\bibinfo{volume}{75}},
  \bibinfo{pages}{6568} (\bibinfo{year}{1994}), ISSN \bibinfo{issn}{0021-8979},
  \urlprefix\url{https://aip.scitation.org/doi/abs/10.1063/1.356923
  http://aip.scitation.org/doi/10.1063/1.356923}.

\bibitem[{\citenamefont{Ching et~al.}(1994)\citenamefont{Ching, Chang, Chin,
  Duh, and Ku}}]{Ching1994}
\bibinfo{author}{\bibfnamefont{K.~M.} \bibnamefont{Ching}},
  \bibinfo{author}{\bibfnamefont{W.~D.} \bibnamefont{Chang}},
  \bibinfo{author}{\bibfnamefont{T.~S.} \bibnamefont{Chin}},
  \bibinfo{author}{\bibfnamefont{J.~G.} \bibnamefont{Duh}}, \bibnamefont{and}
  \bibinfo{author}{\bibfnamefont{H.~C.} \bibnamefont{Ku}},
  \bibinfo{journal}{Journal of Applied Physics} \textbf{\bibinfo{volume}{76}},
  \bibinfo{pages}{6582} (\bibinfo{year}{1994}), ISSN \bibinfo{issn}{0021-8979},
  \urlprefix\url{https://aip.scitation.org/doi/abs/10.1063/1.358200
  http://aip.scitation.org/doi/10.1063/1.358200}.

\bibitem[{\citenamefont{Yasutomi et~al.}(2014)\citenamefont{Yasutomi, Ito,
  Sanai, Toko, and Suemasu}}]{Yasutomi2014}
\bibinfo{author}{\bibfnamefont{Y.}~\bibnamefont{Yasutomi}},
  \bibinfo{author}{\bibfnamefont{K.}~\bibnamefont{Ito}},
  \bibinfo{author}{\bibfnamefont{T.}~\bibnamefont{Sanai}},
  \bibinfo{author}{\bibfnamefont{K.}~\bibnamefont{Toko}}, \bibnamefont{and}
  \bibinfo{author}{\bibfnamefont{T.}~\bibnamefont{Suemasu}},
  \bibinfo{journal}{Journal of Applied Physics} \textbf{\bibinfo{volume}{115}},
  \bibinfo{pages}{17A935} (\bibinfo{year}{2014}), ISSN
  \bibinfo{issn}{0021-8979},
  \urlprefix\url{https://aip.scitation.org/doi/abs/10.1063/1.4867955
  http://aip.scitation.org/doi/10.1063/1.4867955}.

\bibitem[{\citenamefont{Kabara and Tsunoda}(2015)}]{Kabara2015}
\bibinfo{author}{\bibfnamefont{K.}~\bibnamefont{Kabara}} \bibnamefont{and}
  \bibinfo{author}{\bibfnamefont{M.}~\bibnamefont{Tsunoda}},
  \bibinfo{journal}{Journal of Applied Physics} \textbf{\bibinfo{volume}{117}},
  \bibinfo{pages}{17B512} (\bibinfo{year}{2015}), ISSN
  \bibinfo{issn}{0021-8979},
  \urlprefix\url{https://aip.scitation.org/doi/abs/10.1063/1.4913730
  http://aip.scitation.org/doi/10.1063/1.4913730}.

\bibitem[{\citenamefont{Isogami et~al.}(2020)\citenamefont{Isogami, Masuda, and
  Miura}}]{Isogami2020}
\bibinfo{author}{\bibfnamefont{S.}~\bibnamefont{Isogami}},
  \bibinfo{author}{\bibfnamefont{K.}~\bibnamefont{Masuda}}, \bibnamefont{and}
  \bibinfo{author}{\bibfnamefont{Y.}~\bibnamefont{Miura}},
  \bibinfo{journal}{Physical Review Materials} \textbf{\bibinfo{volume}{4}},
  \bibinfo{pages}{014406} (\bibinfo{year}{2020}), ISSN
  \bibinfo{issn}{2475-9953},
  \urlprefix\url{https://journals.aps.org/prmaterials/abstract/10.1103/PhysRevMaterials.4.014406
  https://link.aps.org/doi/10.1103/PhysRevMaterials.4.014406}.

\bibitem[{\citenamefont{Takei et~al.}(1960)\citenamefont{Takei, Shirane, and
  Frazer}}]{Takei1960}
\bibinfo{author}{\bibfnamefont{W.~J.} \bibnamefont{Takei}},
  \bibinfo{author}{\bibfnamefont{G.}~\bibnamefont{Shirane}}, \bibnamefont{and}
  \bibinfo{author}{\bibfnamefont{B.~C.} \bibnamefont{Frazer}},
  \bibinfo{journal}{Physical Review} \textbf{\bibinfo{volume}{119}},
  \bibinfo{pages}{122} (\bibinfo{year}{1960}), ISSN \bibinfo{issn}{0031-899X},
  \urlprefix\url{https://journals.aps.org/pr/abstract/10.1103/PhysRev.119.122
  https://link.aps.org/doi/10.1103/PhysRev.119.122}.

\bibitem[{\citenamefont{Takei et~al.}(1962)\citenamefont{Takei, Heikes, and
  Shirane}}]{Takei1962}
\bibinfo{author}{\bibfnamefont{W.~J.} \bibnamefont{Takei}},
  \bibinfo{author}{\bibfnamefont{R.~R.} \bibnamefont{Heikes}},
  \bibnamefont{and} \bibinfo{author}{\bibfnamefont{G.}~\bibnamefont{Shirane}},
  \bibinfo{journal}{Physical Review} \textbf{\bibinfo{volume}{125}},
  \bibinfo{pages}{1893} (\bibinfo{year}{1962}), ISSN \bibinfo{issn}{0031-899X},
  \urlprefix\url{https://journals.aps.org/pr/abstract/10.1103/PhysRev.125.1893
  https://link.aps.org/doi/10.1103/PhysRev.125.1893}.

\bibitem[{\citenamefont{Gushi et~al.}(2019)\citenamefont{Gushi,
  {Jovi{\v{c}}evi{\'{c}} Klug}, {Pe{\~{n}}a Garcia}, Ghosh, Attan{\'{e}},
  Okuno, Fruchart, Vogel, Suemasu, Pizzini et~al.}}]{Gushi2019}
\bibinfo{author}{\bibfnamefont{T.}~\bibnamefont{Gushi}},
  \bibinfo{author}{\bibfnamefont{M.}~\bibnamefont{{Jovi{\v{c}}evi{\'{c}}
  Klug}}}, \bibinfo{author}{\bibfnamefont{J.}~\bibnamefont{{Pe{\~{n}}a
  Garcia}}}, \bibinfo{author}{\bibfnamefont{S.}~\bibnamefont{Ghosh}},
  \bibinfo{author}{\bibfnamefont{J.~P.} \bibnamefont{Attan{\'{e}}}},
  \bibinfo{author}{\bibfnamefont{H.}~\bibnamefont{Okuno}},
  \bibinfo{author}{\bibfnamefont{O.}~\bibnamefont{Fruchart}},
  \bibinfo{author}{\bibfnamefont{J.}~\bibnamefont{Vogel}},
  \bibinfo{author}{\bibfnamefont{T.}~\bibnamefont{Suemasu}},
  \bibinfo{author}{\bibfnamefont{S.}~\bibnamefont{Pizzini}},
  \bibnamefont{et~al.}, \bibinfo{journal}{Nano Letters}
  \textbf{\bibinfo{volume}{19}}, \bibinfo{pages}{8716} (\bibinfo{year}{2019}),
  ISSN \bibinfo{issn}{15306992},
  \urlprefix\url{https://pubs.acs.org/doi/abs/10.1021/acs.nanolett.9b03416}.

\bibitem[{\citenamefont{Ghosh et~al.}(2021)\citenamefont{Ghosh, Komori, Hallal,
  {Pe{\~{n}}a Garcia}, Gushi, Hirose, Mitarai, Okuno, Vogel, Chshiev
  et~al.}}]{Ghosh2021}
\bibinfo{author}{\bibfnamefont{S.}~\bibnamefont{Ghosh}},
  \bibinfo{author}{\bibfnamefont{T.}~\bibnamefont{Komori}},
  \bibinfo{author}{\bibfnamefont{A.}~\bibnamefont{Hallal}},
  \bibinfo{author}{\bibfnamefont{J.}~\bibnamefont{{Pe{\~{n}}a Garcia}}},
  \bibinfo{author}{\bibfnamefont{T.}~\bibnamefont{Gushi}},
  \bibinfo{author}{\bibfnamefont{T.}~\bibnamefont{Hirose}},
  \bibinfo{author}{\bibfnamefont{H.}~\bibnamefont{Mitarai}},
  \bibinfo{author}{\bibfnamefont{H.}~\bibnamefont{Okuno}},
  \bibinfo{author}{\bibfnamefont{J.}~\bibnamefont{Vogel}},
  \bibinfo{author}{\bibfnamefont{M.}~\bibnamefont{Chshiev}},
  \bibnamefont{et~al.}, \bibinfo{journal}{Nano Letters}
  \textbf{\bibinfo{volume}{21}}, \bibinfo{pages}{2580} (\bibinfo{year}{2021}),
  ISSN \bibinfo{issn}{15306992}, \eprint{2101.04220},
  \urlprefix\url{https://pubs.acs.org/doi/abs/10.1021/acs.nanolett.1c00125}.

\bibitem[{\citenamefont{Isogami
  et~al.}(2021{\natexlab{a}})\citenamefont{Isogami, Rajamanickam, Kozuka, and
  Takahashi}}]{Isogami2021a}
\bibinfo{author}{\bibfnamefont{S.}~\bibnamefont{Isogami}},
  \bibinfo{author}{\bibfnamefont{N.}~\bibnamefont{Rajamanickam}},
  \bibinfo{author}{\bibfnamefont{Y.}~\bibnamefont{Kozuka}}, \bibnamefont{and}
  \bibinfo{author}{\bibfnamefont{Y.~K.} \bibnamefont{Takahashi}},
  \bibinfo{journal}{AIP Advances} \textbf{\bibinfo{volume}{11}},
  \bibinfo{pages}{105314} (\bibinfo{year}{2021}{\natexlab{a}}), ISSN
  \bibinfo{issn}{2158-3226},
  \urlprefix\url{https://aip.scitation.org/doi/abs/10.1063/5.0062253
  https://aip.scitation.org/doi/10.1063/5.0062253}.

\bibitem[{\citenamefont{Isogami
  et~al.}(2021{\natexlab{b}})\citenamefont{Isogami, Masuda, Miura,
  Rajamanickam, and Sakuraba}}]{Isogami2021b}
\bibinfo{author}{\bibfnamefont{S.}~\bibnamefont{Isogami}},
  \bibinfo{author}{\bibfnamefont{K.}~\bibnamefont{Masuda}},
  \bibinfo{author}{\bibfnamefont{Y.}~\bibnamefont{Miura}},
  \bibinfo{author}{\bibfnamefont{N.}~\bibnamefont{Rajamanickam}},
  \bibnamefont{and} \bibinfo{author}{\bibfnamefont{Y.}~\bibnamefont{Sakuraba}},
  \bibinfo{journal}{Applied Physics Letters} \textbf{\bibinfo{volume}{118}},
  \bibinfo{pages}{092407} (\bibinfo{year}{2021}{\natexlab{b}}), ISSN
  \bibinfo{issn}{0003-6951},
  \urlprefix\url{https://aip.scitation.org/doi/abs/10.1063/5.0039569
  https://aip.scitation.org/doi/10.1063/5.0039569}.

\bibitem[{\citenamefont{Isogami
  et~al.}(2022{\natexlab{a}})\citenamefont{Isogami, Ohtake, and
  Takahashi}}]{Isogami2022}
\bibinfo{author}{\bibfnamefont{S.}~\bibnamefont{Isogami}},
  \bibinfo{author}{\bibfnamefont{M.}~\bibnamefont{Ohtake}}, \bibnamefont{and}
  \bibinfo{author}{\bibfnamefont{Y.~K.} \bibnamefont{Takahashi}},
  \bibinfo{journal}{Journal of Applied Physics} \textbf{\bibinfo{volume}{131}},
  \bibinfo{pages}{073904} (\bibinfo{year}{2022}{\natexlab{a}}), ISSN
  \bibinfo{issn}{0021-8979},
  \urlprefix\url{https://aip.scitation.org/doi/abs/10.1063/5.0083042
  https://aip.scitation.org/doi/10.1063/5.0083042}.

\bibitem[{\citenamefont{Isogami
  et~al.}(2022{\natexlab{b}})\citenamefont{Isogami, Ohtake, Kozuka, and
  Takahashi}}]{Isogami2022a}
\bibinfo{author}{\bibfnamefont{S.}~\bibnamefont{Isogami}},
  \bibinfo{author}{\bibfnamefont{M.}~\bibnamefont{Ohtake}},
  \bibinfo{author}{\bibfnamefont{Y.}~\bibnamefont{Kozuka}}, \bibnamefont{and}
  \bibinfo{author}{\bibfnamefont{Y.~K.} \bibnamefont{Takahashi}},
  \bibinfo{journal}{Journal of Magnetism and Magnetic Materials}
  \textbf{\bibinfo{volume}{560}}, \bibinfo{pages}{169642}
  (\bibinfo{year}{2022}{\natexlab{b}}), ISSN \bibinfo{issn}{0304-8853}.

\bibitem[{\citenamefont{Chen et~al.}(2022)\citenamefont{Chen, Shi, Liu, Chen,
  Zhang, and Mi}}]{Chen2022}
\bibinfo{author}{\bibfnamefont{Z.}~\bibnamefont{Chen}},
  \bibinfo{author}{\bibfnamefont{X.}~\bibnamefont{Shi}},
  \bibinfo{author}{\bibfnamefont{X.}~\bibnamefont{Liu}},
  \bibinfo{author}{\bibfnamefont{X.}~\bibnamefont{Chen}},
  \bibinfo{author}{\bibfnamefont{Z.}~\bibnamefont{Zhang}}, \bibnamefont{and}
  \bibinfo{author}{\bibfnamefont{W.}~\bibnamefont{Mi}},
  \bibinfo{journal}{Journal of Applied Physics} \textbf{\bibinfo{volume}{132}},
  \bibinfo{pages}{233906} (\bibinfo{year}{2022}), ISSN
  \bibinfo{issn}{0021-8979},
  \urlprefix\url{https://aip.scitation.org/doi/abs/10.1063/5.0133067}.

\bibitem[{\citenamefont{Zhang et~al.}(2022{\natexlab{a}})\citenamefont{Zhang,
  Zhang, and Mi}}]{Zhang2022a}
\bibinfo{author}{\bibfnamefont{Z.}~\bibnamefont{Zhang}},
  \bibinfo{author}{\bibfnamefont{Q.}~\bibnamefont{Zhang}}, \bibnamefont{and}
  \bibinfo{author}{\bibfnamefont{W.}~\bibnamefont{Mi}},
  \bibinfo{journal}{Chinese Physics B} \textbf{\bibinfo{volume}{31}},
  \bibinfo{pages}{047305} (\bibinfo{year}{2022}{\natexlab{a}}), ISSN
  \bibinfo{issn}{1674-1056},
  \urlprefix\url{https://iopscience.iop.org/article/10.1088/1674-1056/ac2b90
  https://iopscience.iop.org/article/10.1088/1674-1056/ac2b90/meta}.

\bibitem[{\citenamefont{Zhang et~al.}(2022{\natexlab{b}})\citenamefont{Zhang,
  Jiang, Shi, Liu, Chen, Hou, and Mi}}]{Zhang2022b}
\bibinfo{author}{\bibfnamefont{Z.~Y.} \bibnamefont{Zhang}},
  \bibinfo{author}{\bibfnamefont{J.~W.} \bibnamefont{Jiang}},
  \bibinfo{author}{\bibfnamefont{X.~H.} \bibnamefont{Shi}},
  \bibinfo{author}{\bibfnamefont{X.}~\bibnamefont{Liu}},
  \bibinfo{author}{\bibfnamefont{X.}~\bibnamefont{Chen}},
  \bibinfo{author}{\bibfnamefont{Z.~P.} \bibnamefont{Hou}}, \bibnamefont{and}
  \bibinfo{author}{\bibfnamefont{W.~B.} \bibnamefont{Mi}},
  \bibinfo{journal}{Rare Metals} pp. \bibinfo{pages}{1--11}
  (\bibinfo{year}{2022}{\natexlab{b}}), ISSN \bibinfo{issn}{18677185},
  \urlprefix\url{https://link.springer.com/article/10.1007/s12598-022-02166-z}.

\bibitem[{\citenamefont{Komori et~al.}(2019)\citenamefont{Komori, Gushi, Anzai,
  Vila, Attan{\'{e}}, Pizzini, Vogel, Isogami, Toko, and Suemasu}}]{Komori2019}
\bibinfo{author}{\bibfnamefont{T.}~\bibnamefont{Komori}},
  \bibinfo{author}{\bibfnamefont{T.}~\bibnamefont{Gushi}},
  \bibinfo{author}{\bibfnamefont{A.}~\bibnamefont{Anzai}},
  \bibinfo{author}{\bibfnamefont{L.}~\bibnamefont{Vila}},
  \bibinfo{author}{\bibfnamefont{J.~P.} \bibnamefont{Attan{\'{e}}}},
  \bibinfo{author}{\bibfnamefont{S.}~\bibnamefont{Pizzini}},
  \bibinfo{author}{\bibfnamefont{J.}~\bibnamefont{Vogel}},
  \bibinfo{author}{\bibfnamefont{S.}~\bibnamefont{Isogami}},
  \bibinfo{author}{\bibfnamefont{K.}~\bibnamefont{Toko}}, \bibnamefont{and}
  \bibinfo{author}{\bibfnamefont{T.}~\bibnamefont{Suemasu}},
  \bibinfo{journal}{Journal of Applied Physics} \textbf{\bibinfo{volume}{125}},
  \bibinfo{pages}{213902} (\bibinfo{year}{2019}), ISSN
  \bibinfo{issn}{0021-8979},
  \urlprefix\url{https://aip.scitation.org/doi/abs/10.1063/1.5089869}.

\bibitem[{\citenamefont{Kabara et~al.}(2017)\citenamefont{Kabara, Tsunoda, and
  Kokado}}]{Kabara2017a}
\bibinfo{author}{\bibfnamefont{K.}~\bibnamefont{Kabara}},
  \bibinfo{author}{\bibfnamefont{M.}~\bibnamefont{Tsunoda}}, \bibnamefont{and}
  \bibinfo{author}{\bibfnamefont{S.}~\bibnamefont{Kokado}},
  \bibinfo{journal}{AIP Advances} \textbf{\bibinfo{volume}{7}},
  \bibinfo{pages}{056416} (\bibinfo{year}{2017}), ISSN
  \bibinfo{issn}{21583226},
  \urlprefix\url{https://aip.scitation.org/doi/abs/10.1063/1.4974065}.

\bibitem[{\citenamefont{Ma et~al.}(2021)\citenamefont{Ma, Hartnett, Zhou,
  Balachandran, and Poon}}]{Ma2021}
\bibinfo{author}{\bibfnamefont{C.~T.} \bibnamefont{Ma}},
  \bibinfo{author}{\bibfnamefont{T.~Q.} \bibnamefont{Hartnett}},
  \bibinfo{author}{\bibfnamefont{W.}~\bibnamefont{Zhou}},
  \bibinfo{author}{\bibfnamefont{P.~V.} \bibnamefont{Balachandran}},
  \bibnamefont{and} \bibinfo{author}{\bibfnamefont{S.~J.} \bibnamefont{Poon}},
  \bibinfo{journal}{Applied Physics Letters} \textbf{\bibinfo{volume}{119}},
  \bibinfo{pages}{192406} (\bibinfo{year}{2021}), ISSN
  \bibinfo{issn}{0003-6951},
  \urlprefix\url{https://aip.scitation.org/doi/abs/10.1063/5.0066375
  https://aip.scitation.org/doi/10.1063/5.0066375}.

\bibitem[{\citenamefont{Wang et~al.}(2018)\citenamefont{Wang, Xu, Lou, Liu, Li,
  Huang, Shen, Weng, Wang, and Lei}}]{Wang2018}
\bibinfo{author}{\bibfnamefont{Q.}~\bibnamefont{Wang}},
  \bibinfo{author}{\bibfnamefont{Y.}~\bibnamefont{Xu}},
  \bibinfo{author}{\bibfnamefont{R.}~\bibnamefont{Lou}},
  \bibinfo{author}{\bibfnamefont{Z.}~\bibnamefont{Liu}},
  \bibinfo{author}{\bibfnamefont{M.}~\bibnamefont{Li}},
  \bibinfo{author}{\bibfnamefont{Y.}~\bibnamefont{Huang}},
  \bibinfo{author}{\bibfnamefont{D.}~\bibnamefont{Shen}},
  \bibinfo{author}{\bibfnamefont{H.}~\bibnamefont{Weng}},
  \bibinfo{author}{\bibfnamefont{S.}~\bibnamefont{Wang}}, \bibnamefont{and}
  \bibinfo{author}{\bibfnamefont{H.}~\bibnamefont{Lei}},
  \bibinfo{journal}{Nature Communications 2018 9:1}
  \textbf{\bibinfo{volume}{9}}, \bibinfo{pages}{1} (\bibinfo{year}{2018}), ISSN
  \bibinfo{issn}{2041-1723}, \eprint{1712.09947},
  \urlprefix\url{https://www.nature.com/articles/s41467-018-06088-2}.

\bibitem[{\citenamefont{Bayaraa et~al.}(2021)\citenamefont{Bayaraa, Xu, and
  Bellaiche}}]{Bayaraa2021a}
\bibinfo{author}{\bibfnamefont{T.}~\bibnamefont{Bayaraa}},
  \bibinfo{author}{\bibfnamefont{C.}~\bibnamefont{Xu}}, \bibnamefont{and}
  \bibinfo{author}{\bibfnamefont{L.}~\bibnamefont{Bellaiche}},
  \bibinfo{journal}{Physical Review Letters} \textbf{\bibinfo{volume}{127}},
  \bibinfo{pages}{217204} (\bibinfo{year}{2021}), ISSN
  \bibinfo{issn}{0031-9007},
  \urlprefix\url{https://journals.aps.org/prl/abstract/10.1103/PhysRevLett.127.217204
  https://link.aps.org/doi/10.1103/PhysRevLett.127.217204}.

\bibitem[{\citenamefont{Shen et~al.}(2014)\citenamefont{Shen, Chikamatsu,
  Shigematsu, Hirose, Fukumura, and Hasegawa}}]{Shen2014}
\bibinfo{author}{\bibfnamefont{X.}~\bibnamefont{Shen}},
  \bibinfo{author}{\bibfnamefont{A.}~\bibnamefont{Chikamatsu}},
  \bibinfo{author}{\bibfnamefont{K.}~\bibnamefont{Shigematsu}},
  \bibinfo{author}{\bibfnamefont{Y.}~\bibnamefont{Hirose}},
  \bibinfo{author}{\bibfnamefont{T.}~\bibnamefont{Fukumura}}, \bibnamefont{and}
  \bibinfo{author}{\bibfnamefont{T.}~\bibnamefont{Hasegawa}},
  \bibinfo{journal}{Applied Physics Letters} \textbf{\bibinfo{volume}{105}},
  \bibinfo{pages}{072410} (\bibinfo{year}{2014}), ISSN
  \bibinfo{issn}{0003-6951},
  \urlprefix\url{http://aip.scitation.org/doi/10.1063/1.4893732}.

\bibitem[{\citenamefont{Karplus and Luttinger}(1954)}]{Karplus1954}
\bibinfo{author}{\bibfnamefont{R.}~\bibnamefont{Karplus}} \bibnamefont{and}
  \bibinfo{author}{\bibfnamefont{J.~M.} \bibnamefont{Luttinger}},
  \bibinfo{journal}{Physical Review} \textbf{\bibinfo{volume}{95}},
  \bibinfo{pages}{1154} (\bibinfo{year}{1954}), ISSN \bibinfo{issn}{0031899X},
  \urlprefix\url{https://journals.aps.org/pr/abstract/10.1103/PhysRev.95.1154}.

\bibitem[{\citenamefont{Berger}(1970)}]{Berger1970}
\bibinfo{author}{\bibfnamefont{L.}~\bibnamefont{Berger}},
  \bibinfo{journal}{Physical Review B} \textbf{\bibinfo{volume}{2}},
  \bibinfo{pages}{4559} (\bibinfo{year}{1970}), ISSN \bibinfo{issn}{01631829},
  \urlprefix\url{https://journals.aps.org/prb/abstract/10.1103/PhysRevB.2.4559}.

\bibitem[{\citenamefont{Ito et~al.}(2016)\citenamefont{Ito, Yasutomi, Kabara,
  Gushi, Higashikozono, Toko, Tsunoda, and Suemasu}}]{Ito2016}
\bibinfo{author}{\bibfnamefont{K.}~\bibnamefont{Ito}},
  \bibinfo{author}{\bibfnamefont{Y.}~\bibnamefont{Yasutomi}},
  \bibinfo{author}{\bibfnamefont{K.}~\bibnamefont{Kabara}},
  \bibinfo{author}{\bibfnamefont{T.}~\bibnamefont{Gushi}},
  \bibinfo{author}{\bibfnamefont{S.}~\bibnamefont{Higashikozono}},
  \bibinfo{author}{\bibfnamefont{K.}~\bibnamefont{Toko}},
  \bibinfo{author}{\bibfnamefont{M.}~\bibnamefont{Tsunoda}}, \bibnamefont{and}
  \bibinfo{author}{\bibfnamefont{T.}~\bibnamefont{Suemasu}},
  \bibinfo{journal}{AIP Advances} \textbf{\bibinfo{volume}{6}},
  \bibinfo{pages}{056201} (\bibinfo{year}{2016}), ISSN
  \bibinfo{issn}{2158-3226},
  \urlprefix\url{http://aip.scitation.org/doi/10.1063/1.4942548}.

\bibitem[{\citenamefont{Kresse and Joubert}(1999)}]{Kresse1999}
\bibinfo{author}{\bibfnamefont{G.}~\bibnamefont{Kresse}} \bibnamefont{and}
  \bibinfo{author}{\bibfnamefont{D.}~\bibnamefont{Joubert}},
  \bibinfo{journal}{Physical Review B} \textbf{\bibinfo{volume}{59}},
  \bibinfo{pages}{1758} (\bibinfo{year}{1999}), ISSN \bibinfo{issn}{0163-1829},
  \urlprefix\url{https://link.aps.org/doi/10.1103/PhysRevB.59.1758}.

\bibitem[{\citenamefont{Bl{\"{o}}chl}(1994)}]{Blochl1994}
\bibinfo{author}{\bibfnamefont{P.~E.} \bibnamefont{Bl{\"{o}}chl}},
  \bibinfo{journal}{Physical Review B} \textbf{\bibinfo{volume}{50}},
  \bibinfo{pages}{17953} (\bibinfo{year}{1994}), ISSN
  \bibinfo{issn}{0163-1829},
  \urlprefix\url{https://link.aps.org/doi/10.1103/PhysRevB.50.17953}.

\bibitem[{\citenamefont{Liechtenstein et~al.}(1995)\citenamefont{Liechtenstein,
  Anisimov, and Zaanen}}]{Liechtenstein1995}
\bibinfo{author}{\bibfnamefont{A.~I.} \bibnamefont{Liechtenstein}},
  \bibinfo{author}{\bibfnamefont{V.~I.} \bibnamefont{Anisimov}},
  \bibnamefont{and} \bibinfo{author}{\bibfnamefont{J.}~\bibnamefont{Zaanen}},
  \bibinfo{journal}{Physical Review B} \textbf{\bibinfo{volume}{52}},
  \bibinfo{pages}{R5467} (\bibinfo{year}{1995}), ISSN
  \bibinfo{issn}{0163-1829},
  \urlprefix\url{https://link.aps.org/doi/10.1103/PhysRevB.52.R5467}.

\bibitem[{\citenamefont{Perdew et~al.}(1996)\citenamefont{Perdew, Burke, and
  Ernzerhof}}]{GGA}
\bibinfo{author}{\bibfnamefont{J.~P.} \bibnamefont{Perdew}},
  \bibinfo{author}{\bibfnamefont{K.}~\bibnamefont{Burke}}, \bibnamefont{and}
  \bibinfo{author}{\bibfnamefont{M.}~\bibnamefont{Ernzerhof}},
  \bibinfo{journal}{Physical Review Letters} \textbf{\bibinfo{volume}{77}},
  \bibinfo{pages}{3865} (\bibinfo{year}{1996}), ISSN \bibinfo{issn}{0031-9007},
  \urlprefix\url{https://link.aps.org/doi/10.1103/PhysRevLett.77.3865}.

\bibitem[{\citenamefont{Pizzi et~al.}(2020)\citenamefont{Pizzi, Vitale, Arita,
  Bl{\"{u}}gel, Freimuth, G{\'{e}}ranton, Gibertini, Gresch, Johnson, Koretsune
  et~al.}}]{Wannier90}
\bibinfo{author}{\bibfnamefont{G.}~\bibnamefont{Pizzi}},
  \bibinfo{author}{\bibfnamefont{V.}~\bibnamefont{Vitale}},
  \bibinfo{author}{\bibfnamefont{R.}~\bibnamefont{Arita}},
  \bibinfo{author}{\bibfnamefont{S.}~\bibnamefont{Bl{\"{u}}gel}},
  \bibinfo{author}{\bibfnamefont{F.}~\bibnamefont{Freimuth}},
  \bibinfo{author}{\bibfnamefont{G.}~\bibnamefont{G{\'{e}}ranton}},
  \bibinfo{author}{\bibfnamefont{M.}~\bibnamefont{Gibertini}},
  \bibinfo{author}{\bibfnamefont{D.}~\bibnamefont{Gresch}},
  \bibinfo{author}{\bibfnamefont{C.}~\bibnamefont{Johnson}},
  \bibinfo{author}{\bibfnamefont{T.}~\bibnamefont{Koretsune}},
  \bibnamefont{et~al.}, \bibinfo{journal}{Journal of Physics: Condensed Matter}
  \textbf{\bibinfo{volume}{32}}, \bibinfo{pages}{165902}
  (\bibinfo{year}{2020}), ISSN \bibinfo{issn}{0953-8984}, \eprint{1907.09788},
  \urlprefix\url{https://iopscience.iop.org/article/10.1088/1361-648X/ab51ff
  https://iopscience.iop.org/article/10.1088/1361-648X/ab51ff/meta}.

\bibitem[{\citenamefont{Xiao et~al.}(2006)\citenamefont{Xiao, Yao, Fang, and
  Niu}}]{ANE}
\bibinfo{author}{\bibfnamefont{D.}~\bibnamefont{Xiao}},
  \bibinfo{author}{\bibfnamefont{Y.}~\bibnamefont{Yao}},
  \bibinfo{author}{\bibfnamefont{Z.}~\bibnamefont{Fang}}, \bibnamefont{and}
  \bibinfo{author}{\bibfnamefont{Q.}~\bibnamefont{Niu}},
  \bibinfo{journal}{Physical Review Letters} \textbf{\bibinfo{volume}{97}},
  \bibinfo{pages}{026603} (\bibinfo{year}{2006}), ISSN
  \bibinfo{issn}{00319007}, \eprint{0604561},
  \urlprefix\url{https://journals.aps.org/prl/abstract/10.1103/PhysRevLett.97.026603}.

\bibitem[{\citenamefont{Wu et~al.}(2018)\citenamefont{Wu, Zhang, Song, Troyer,
  and Soluyanov}}]{Wanniertools}
\bibinfo{author}{\bibfnamefont{Q.~S.} \bibnamefont{Wu}},
  \bibinfo{author}{\bibfnamefont{S.~N.} \bibnamefont{Zhang}},
  \bibinfo{author}{\bibfnamefont{H.~F.} \bibnamefont{Song}},
  \bibinfo{author}{\bibfnamefont{M.}~\bibnamefont{Troyer}}, \bibnamefont{and}
  \bibinfo{author}{\bibfnamefont{A.~A.} \bibnamefont{Soluyanov}},
  \bibinfo{journal}{Computer Physics Communications}
  \textbf{\bibinfo{volume}{224}}, \bibinfo{pages}{405} (\bibinfo{year}{2018}),
  ISSN \bibinfo{issn}{0010-4655}, \eprint{1703.07789}.

\bibitem[{\citenamefont{Stokes and Hatch}(2005)}]{Stokes2005}
\bibinfo{author}{\bibfnamefont{H.~T.} \bibnamefont{Stokes}} \bibnamefont{and}
  \bibinfo{author}{\bibfnamefont{D.~M.} \bibnamefont{Hatch}},
  \bibinfo{journal}{J. Appl. Cryst. J. Appl. Cryst}
  \textbf{\bibinfo{volume}{38}}, \bibinfo{pages}{237} (\bibinfo{year}{2005}),
  ISSN \bibinfo{issn}{0021-8898},
  \urlprefix\url{http://journals.iucr.org/j/services/authorservices.html.}

\bibitem[{\citenamefont{Komori et~al.}(2020)\citenamefont{Komori, Hirose,
  Gushi, Toko, Hanashima, Vila, Attane, Amemiya, and Suemasu}}]{Komori2020}
\bibinfo{author}{\bibfnamefont{T.}~\bibnamefont{Komori}},
  \bibinfo{author}{\bibfnamefont{T.}~\bibnamefont{Hirose}},
  \bibinfo{author}{\bibfnamefont{T.}~\bibnamefont{Gushi}},
  \bibinfo{author}{\bibfnamefont{K.}~\bibnamefont{Toko}},
  \bibinfo{author}{\bibfnamefont{Î.}~\bibnamefont{Hanashima}},
  \bibinfo{author}{\bibfnamefont{L.}~\bibnamefont{Vila}},
  \bibinfo{author}{\bibfnamefont{J.-P.} \bibnamefont{Attane}},
  \bibinfo{author}{\bibfnamefont{K.}~\bibnamefont{Amemiya}}, \bibnamefont{and}
  \bibinfo{author}{\bibfnamefont{T.}~\bibnamefont{Suemasu}},
  \bibinfo{journal}{Journal of Applied Physics} \textbf{\bibinfo{volume}{127}},
  \bibinfo{pages}{043903} (\bibinfo{year}{2020}), ISSN
  \bibinfo{issn}{0021-8979},
  \urlprefix\url{https://aip.scitation.org/doi/abs/10.1063/1.5128635
  http://aip.scitation.org/doi/10.1063/1.5128635}.

\bibitem[{\citenamefont{Komori et~al.}(2022{\natexlab{a}})\citenamefont{Komori,
  Mitarai, Yasuda, Ghosh, Vila, Attan{\'{e}}, Honda, and
  Suemasu}}]{Komori2022a}
\bibinfo{author}{\bibfnamefont{T.}~\bibnamefont{Komori}},
  \bibinfo{author}{\bibfnamefont{H.}~\bibnamefont{Mitarai}},
  \bibinfo{author}{\bibfnamefont{T.}~\bibnamefont{Yasuda}},
  \bibinfo{author}{\bibfnamefont{S.}~\bibnamefont{Ghosh}},
  \bibinfo{author}{\bibfnamefont{L.}~\bibnamefont{Vila}},
  \bibinfo{author}{\bibfnamefont{J.~P.} \bibnamefont{Attan{\'{e}}}},
  \bibinfo{author}{\bibfnamefont{S.}~\bibnamefont{Honda}}, \bibnamefont{and}
  \bibinfo{author}{\bibfnamefont{T.}~\bibnamefont{Suemasu}},
  \bibinfo{journal}{Journal of Applied Physics} \textbf{\bibinfo{volume}{132}},
  \bibinfo{pages}{143902} (\bibinfo{year}{2022}{\natexlab{a}}), ISSN
  \bibinfo{issn}{0021-8979},
  \urlprefix\url{https://aip.scitation.org/doi/abs/10.1063/5.0107172}.

\bibitem[{\citenamefont{Komori et~al.}(2022{\natexlab{b}})\citenamefont{Komori,
  Horiuchi, Mitarai, Yasuda, Amemiya, and Suemasu}}]{Komori2022b}
\bibinfo{author}{\bibfnamefont{T.}~\bibnamefont{Komori}},
  \bibinfo{author}{\bibfnamefont{T.}~\bibnamefont{Horiuchi}},
  \bibinfo{author}{\bibfnamefont{H.}~\bibnamefont{Mitarai}},
  \bibinfo{author}{\bibfnamefont{T.}~\bibnamefont{Yasuda}},
  \bibinfo{author}{\bibfnamefont{K.}~\bibnamefont{Amemiya}}, \bibnamefont{and}
  \bibinfo{author}{\bibfnamefont{T.}~\bibnamefont{Suemasu}},
  \bibinfo{journal}{Journal of Magnetism and Magnetic Materials}
  \textbf{\bibinfo{volume}{564}}, \bibinfo{pages}{170050}
  (\bibinfo{year}{2022}{\natexlab{b}}), ISSN \bibinfo{issn}{0304-8853}.

\bibitem[{\citenamefont{Mitarai et~al.}(2020)\citenamefont{Mitarai, Komori,
  Hirose, Ito, Ghosh, Honda, Toko, Vila, Attan{\'{e}}, Amemiya
  et~al.}}]{Mitarai2020}
\bibinfo{author}{\bibfnamefont{H.}~\bibnamefont{Mitarai}},
  \bibinfo{author}{\bibfnamefont{T.}~\bibnamefont{Komori}},
  \bibinfo{author}{\bibfnamefont{T.}~\bibnamefont{Hirose}},
  \bibinfo{author}{\bibfnamefont{K.}~\bibnamefont{Ito}},
  \bibinfo{author}{\bibfnamefont{S.}~\bibnamefont{Ghosh}},
  \bibinfo{author}{\bibfnamefont{S.}~\bibnamefont{Honda}},
  \bibinfo{author}{\bibfnamefont{K.}~\bibnamefont{Toko}},
  \bibinfo{author}{\bibfnamefont{L.}~\bibnamefont{Vila}},
  \bibinfo{author}{\bibfnamefont{J.-P.} \bibnamefont{Attan{\'{e}}}},
  \bibinfo{author}{\bibfnamefont{K.}~\bibnamefont{Amemiya}},
  \bibnamefont{et~al.}, \bibinfo{journal}{Physical Review Materials}
  \textbf{\bibinfo{volume}{4}}, \bibinfo{pages}{094401} (\bibinfo{year}{2020}),
  ISSN \bibinfo{issn}{2475-9953},
  \urlprefix\url{https://journals.aps.org/prmaterials/abstract/10.1103/PhysRevMaterials.4.094401
  https://link.aps.org/doi/10.1103/PhysRevMaterials.4.094401}.

\bibitem[{\citenamefont{Asaba et~al.}(2021)\citenamefont{Asaba, Ivanov, Thomas,
  Savrasov, Thompson, Bauer, and Ronning}}]{Asaba2021}
\bibinfo{author}{\bibfnamefont{T.}~\bibnamefont{Asaba}},
  \bibinfo{author}{\bibfnamefont{V.}~\bibnamefont{Ivanov}},
  \bibinfo{author}{\bibfnamefont{S.~M.} \bibnamefont{Thomas}},
  \bibinfo{author}{\bibfnamefont{S.~Y.} \bibnamefont{Savrasov}},
  \bibinfo{author}{\bibfnamefont{J.~D.} \bibnamefont{Thompson}},
  \bibinfo{author}{\bibfnamefont{E.~D.} \bibnamefont{Bauer}}, \bibnamefont{and}
  \bibinfo{author}{\bibfnamefont{F.}~\bibnamefont{Ronning}},
  \bibinfo{journal}{Science Advances} \textbf{\bibinfo{volume}{7}}
  (\bibinfo{year}{2021}), ISSN \bibinfo{issn}{23752548}, \eprint{2104.09060},
  \urlprefix\url{https://www.science.org/doi/10.1126/sciadv.abf1467}.

\bibitem[{\citenamefont{Yang et~al.}(2020)\citenamefont{Yang, You, Wang, Huang,
  Xi, Xu, Cao, Tian, Xu, Dai et~al.}}]{Yang2020}
\bibinfo{author}{\bibfnamefont{H.}~\bibnamefont{Yang}},
  \bibinfo{author}{\bibfnamefont{W.}~\bibnamefont{You}},
  \bibinfo{author}{\bibfnamefont{J.}~\bibnamefont{Wang}},
  \bibinfo{author}{\bibfnamefont{J.}~\bibnamefont{Huang}},
  \bibinfo{author}{\bibfnamefont{C.}~\bibnamefont{Xi}},
  \bibinfo{author}{\bibfnamefont{X.}~\bibnamefont{Xu}},
  \bibinfo{author}{\bibfnamefont{C.}~\bibnamefont{Cao}},
  \bibinfo{author}{\bibfnamefont{M.}~\bibnamefont{Tian}},
  \bibinfo{author}{\bibfnamefont{Z.~A.} \bibnamefont{Xu}},
  \bibinfo{author}{\bibfnamefont{J.}~\bibnamefont{Dai}}, \bibnamefont{et~al.},
  \bibinfo{journal}{Physical Review Materials} \textbf{\bibinfo{volume}{4}},
  \bibinfo{pages}{024202} (\bibinfo{year}{2020}), ISSN
  \bibinfo{issn}{24759953}, \eprint{1811.03485},
  \urlprefix\url{https://journals.aps.org/prmaterials/abstract/10.1103/PhysRevMaterials.4.024202}.

\bibitem[{\citenamefont{Sakai et~al.}(2018)\citenamefont{Sakai, Mizuta,
  Nugroho, Sihombing, Koretsune, Suzuki, Takemori, Ishii, Nishio-Hamane, Arita
  et~al.}}]{Sakai2022}
\bibinfo{author}{\bibfnamefont{A.}~\bibnamefont{Sakai}},
  \bibinfo{author}{\bibfnamefont{Y.~P.} \bibnamefont{Mizuta}},
  \bibinfo{author}{\bibfnamefont{A.~A.} \bibnamefont{Nugroho}},
  \bibinfo{author}{\bibfnamefont{R.}~\bibnamefont{Sihombing}},
  \bibinfo{author}{\bibfnamefont{T.}~\bibnamefont{Koretsune}},
  \bibinfo{author}{\bibfnamefont{M.~T.} \bibnamefont{Suzuki}},
  \bibinfo{author}{\bibfnamefont{N.}~\bibnamefont{Takemori}},
  \bibinfo{author}{\bibfnamefont{R.}~\bibnamefont{Ishii}},
  \bibinfo{author}{\bibfnamefont{D.}~\bibnamefont{Nishio-Hamane}},
  \bibinfo{author}{\bibfnamefont{R.}~\bibnamefont{Arita}},
  \bibnamefont{et~al.}, \bibinfo{journal}{Nature Physics 2018 14:11}
  \textbf{\bibinfo{volume}{14}}, \bibinfo{pages}{1119} (\bibinfo{year}{2018}),
  ISSN \bibinfo{issn}{1745-2481}, \eprint{1807.04761},
  \urlprefix\url{https://www.nature.com/articles/s41567-018-0225-6}.

\bibitem[{\citenamefont{Miyasato et~al.}(2007)\citenamefont{Miyasato, Abe,
  Fujii, Asamitsu, Onoda, Onose, Nagaosa, and Tokura}}]{Miyasato2007}
\bibinfo{author}{\bibfnamefont{T.}~\bibnamefont{Miyasato}},
  \bibinfo{author}{\bibfnamefont{N.}~\bibnamefont{Abe}},
  \bibinfo{author}{\bibfnamefont{T.}~\bibnamefont{Fujii}},
  \bibinfo{author}{\bibfnamefont{A.}~\bibnamefont{Asamitsu}},
  \bibinfo{author}{\bibfnamefont{S.}~\bibnamefont{Onoda}},
  \bibinfo{author}{\bibfnamefont{Y.}~\bibnamefont{Onose}},
  \bibinfo{author}{\bibfnamefont{N.}~\bibnamefont{Nagaosa}}, \bibnamefont{and}
  \bibinfo{author}{\bibfnamefont{Y.}~\bibnamefont{Tokura}},
  \bibinfo{journal}{Physical Review Letters} \textbf{\bibinfo{volume}{99}},
  \bibinfo{pages}{086602} (\bibinfo{year}{2007}), ISSN
  \bibinfo{issn}{00319007},
  \urlprefix\url{https://journals.aps.org/prl/abstract/10.1103/PhysRevLett.99.086602}.

\bibitem[{\citenamefont{Ikhlas et~al.}(2017)\citenamefont{Ikhlas, Tomita,
  Koretsune, Suzuki, Nishio-Hamane, Arita, Otani, and Nakatsuji}}]{Ikhlas2017}
\bibinfo{author}{\bibfnamefont{M.}~\bibnamefont{Ikhlas}},
  \bibinfo{author}{\bibfnamefont{T.}~\bibnamefont{Tomita}},
  \bibinfo{author}{\bibfnamefont{T.}~\bibnamefont{Koretsune}},
  \bibinfo{author}{\bibfnamefont{M.~T.} \bibnamefont{Suzuki}},
  \bibinfo{author}{\bibfnamefont{D.}~\bibnamefont{Nishio-Hamane}},
  \bibinfo{author}{\bibfnamefont{R.}~\bibnamefont{Arita}},
  \bibinfo{author}{\bibfnamefont{Y.}~\bibnamefont{Otani}}, \bibnamefont{and}
  \bibinfo{author}{\bibfnamefont{S.}~\bibnamefont{Nakatsuji}},
  \bibinfo{journal}{Nature Physics 2017 13:11} \textbf{\bibinfo{volume}{13}},
  \bibinfo{pages}{1085} (\bibinfo{year}{2017}), ISSN \bibinfo{issn}{1745-2481},
  \eprint{1710.00062},
  \urlprefix\url{https://www.nature.com/articles/nphys4181}.

\bibitem[{\citenamefont{Ivanov et~al.}(2022)\citenamefont{Ivanov, Banyas, and
  Tan}}]{Seva}
\bibinfo{author}{\bibfnamefont{V.}~\bibnamefont{Ivanov}},
  \bibinfo{author}{\bibfnamefont{E.}~\bibnamefont{Banyas}}, \bibnamefont{and}
  \bibinfo{author}{\bibfnamefont{L.~Z.} \bibnamefont{Tan}}
  (\bibinfo{year}{2022}), \eprint{2211.07773},
  \urlprefix\url{https://arxiv.org/abs/2211.07773v1}.

\bibitem[{\citenamefont{Weber et~al.}(2019)\citenamefont{Weber, Griffin, and
  Neaton}}]{Weber2019}
\bibinfo{author}{\bibfnamefont{S.~F.} \bibnamefont{Weber}},
  \bibinfo{author}{\bibfnamefont{S.~M.} \bibnamefont{Griffin}},
  \bibnamefont{and} \bibinfo{author}{\bibfnamefont{J.~B.}
  \bibnamefont{Neaton}}, \bibinfo{journal}{Physical Review Materials}
  \textbf{\bibinfo{volume}{3}}, \bibinfo{pages}{064206} (\bibinfo{year}{2019}),
  ISSN \bibinfo{issn}{24759953}, \eprint{1902.10085},
  \urlprefix\url{https://journals.aps.org/prmaterials/abstract/10.1103/PhysRevMaterials.3.064206}.

\end{thebibliography}

\end{document}

\preprint{APS/123-QED}

\title{Intrinsic Origin and Enhancement of Topological Responses in Ferrimagnetic Antiperovskite Mn$_4$N}
\author{Temuujin Bayaraa$^{1,2}$}
\email{tbayaraa@lbl.gov}
\author{Vsevolod Ivanov$^{1,3}$}
\author{Liang Z. Tan$^2$}
\author{Sinéad M. Griffin$^{1,2}$}
\affiliation{$^{1}$Materials Sciences Division, Lawrence Berkeley National Laboratory, Berkeley, California, 94720, USA}
\affiliation{$^{2}$Molecular Foundry Division, Lawrence Berkeley National Laboratory, Berkeley, California, 94720, USA}
\affiliation{$^{3}$Accelerator Technology and Applied Physics Division, Lawrence Berkeley National Laboratory, Berkeley, California, 94720, USA}

\date{\today}

\begin{abstract}

The purpose of this Supplemental Material is to show identified Weyl points in the Brillouin zone for the bulk and the compressive strain of -3\% cases and provide a more in-depth understanding of the change in orbital projections on the band-structures and their overlap under compressive strain.   

\end{abstract}

\maketitle

Fig.~S1(a) shows the Weyl point (WP) in the Brillouin zone (BZ) that we identified as the source of intense Berry curvature ($\Omega_z$) in Fig. 1(b) of the manuscript for the bulk Mn$_4$N. Fig.~S1(b) shows the WP that derives the negative $\Omega_z$ in Fig. 4(c) of the manuscript for the compressive strain case of -3\%. We find that such WP in the bulk case resides between high-symmetry points \textit{R} and \textit{M}, and WP in the -3\% film resides between \textit{M} and \textit{A}. Other WPs in the BZ exist which are not plotted since they do not contribute as strongly to the Berry curvature as those shown in Fig S1. Moreover, we summarize the symmetry inequivalent locations and energies of the WPs identified in the bulk in Table S1. Note that we considered all the bands crossing the Fermi level along the \textit{$\Gamma$-X-R-M-A-Z} line in the WP identification process. 

\beginsupplement
\begin{table}[]
\vspace{10pt}
\begin{tabular}{|c|c|c|c|}
\hline
 $|k_x|$ & \multicolumn{1}{l|}{ $|k_y|$ } & \multicolumn{1}{l|}{$|k_z|$} & \multicolumn{1}{l|}{Energy (eV)} \\ \hline
        0.337 & 0.337 & 0.082 & -0.442\\
        0.407 & 0.498 & 0.001 & -0.198\\
        0.500 & 0.500 & 0.273 & -0.005\\
        0.441 & 0.323 & 0.270 & 0.322\\
        0.200 & 0.500 & 0.419 & 0.420\\
        0.270 & 0.470 & 0.389 & 0.459\\
        0.000 & 0.432 & 0.019 & 0.552\\
        0.005 & 0.500 & 0.030 & 0.558\\ \hline
    \end{tabular}
    \caption{Cartesian locations in units of 2$\pi$/a and the energies of the identified symmetry-inequivalent WPs in the bulk (unstrained) Mn$_4$N.}
    \label{tab:my_label}
\end{table}

Fig. S2 shows the projections of the Mn-\textit{d} orbitals for each inequivalent Mn along \textit{R-M-A-Z} in an 0.8 eV window surrounding the Fermi level. We find the largest contributions to the WPs and the band crossings near the Fermi level are the d$_{xy}$, d$_{yz}$, d$_{xz}$, and d$_{x^2-y^2}$ of Mn III.  We note that there are two different Mn III ions, located at (0.5, 0, 0.5) and (0, 0.5, 0.5) which results in the interchange of the BZ directional dependence of the $d_{xy}$ and $d_{yz}$ orbitals. 

To understand how compressive strain influences these Mn III \textit{d} orbitals, we plot cartoons of how these orbitals evolve with strain in Fig.~S3. As compressive strain is applied and in-plane lattice constants shrink, we see the biggest effect is on the (in-plane) $d_{xy}$ orbitals, as expected from the locations of the Mn III at edge centers. As shown in Fig.~3 in the manuscript, the d$_{xy}$ orbitals shift upwards in energy to a larger extent than the d$_{yz}$ or d$_{xz}$ because of their greater in-plane orbital overlap (see Fig S3).  Under compressive strain, the d$_{x^2-y^2}$ move closer with anti-bonding character, however, while the d$_{xy}$, d$_{yz}$, d$_{xz}$ move closer, their bonds constructively interfere. Thus, the d$_{x^2-y^2}$ orbitals not change under compressive strain however bands with other d-orbital character shift upwards (see Fig 3. in the manuscript). Note that the Fermi level also changes with respect to its bulk case in the strained cases.

\begin{figure*}
\includegraphics[width=0.8\textwidth]{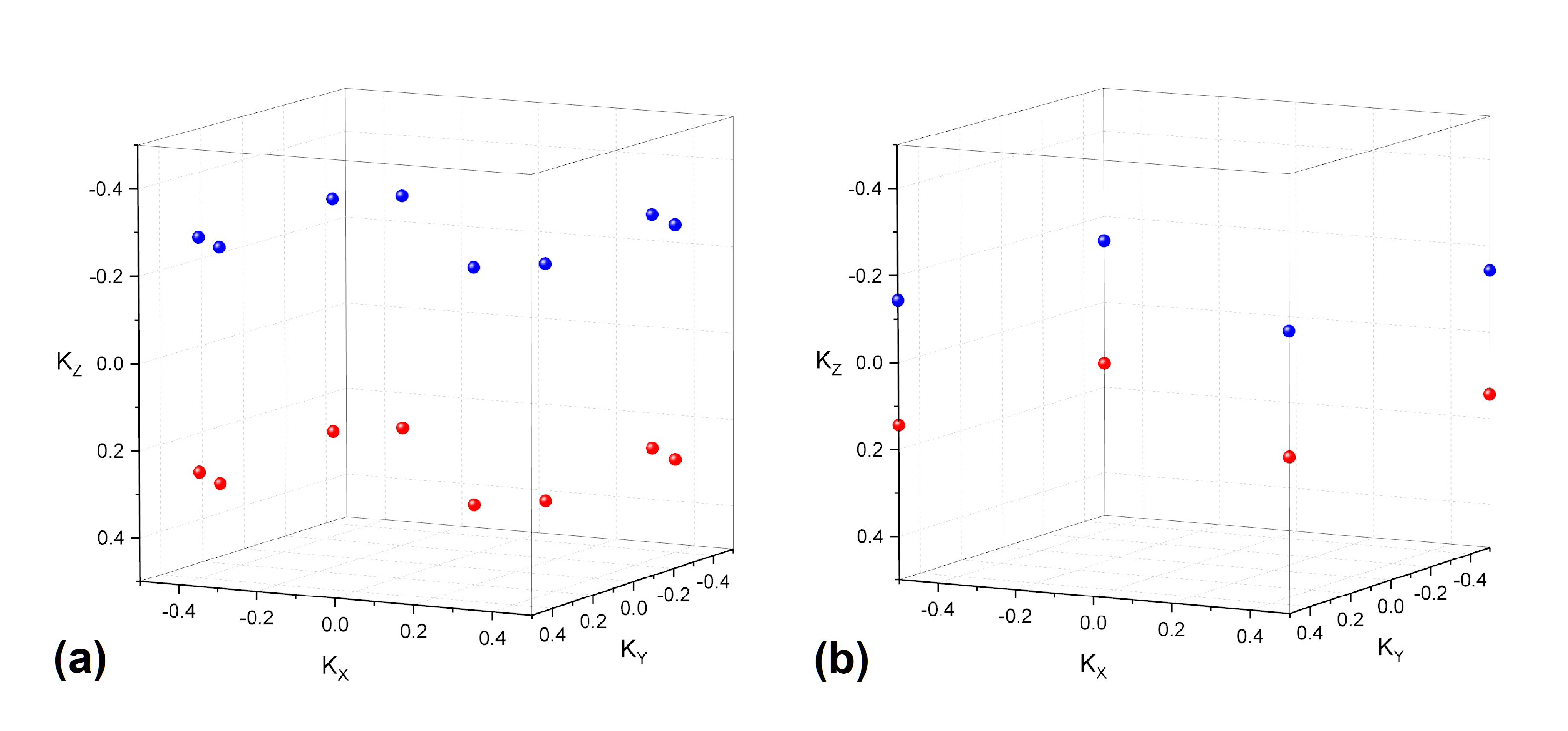}
\caption{Weyl Points that contribute most to the Berry curvature in the (a) bulk and (b) the compressive strain of -3\% cases. Red and blue colors indicate different chiralities of the Weyl points.}
\end{figure*}

\begin{figure*}
\includegraphics[width=0.8\textwidth]{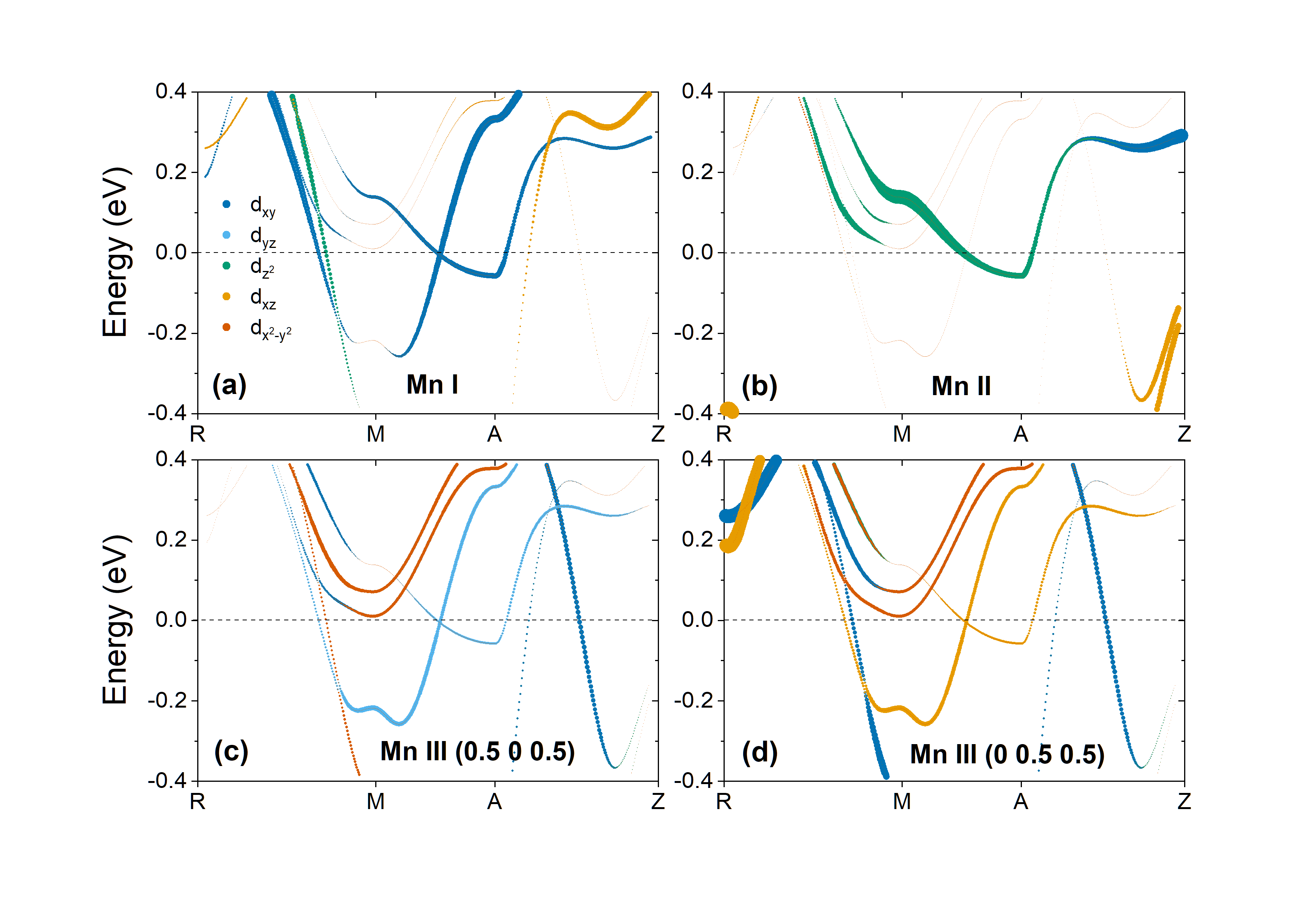}
\caption{d-orbitals- and atom-projected band structures with spin-orbit coupling included for bulk Mn$_4$N over a selected path in the BZ. The Fermi level is set to 0 eV.}
\end{figure*}

\begin{figure*}
\includegraphics[width=0.5\textwidth]{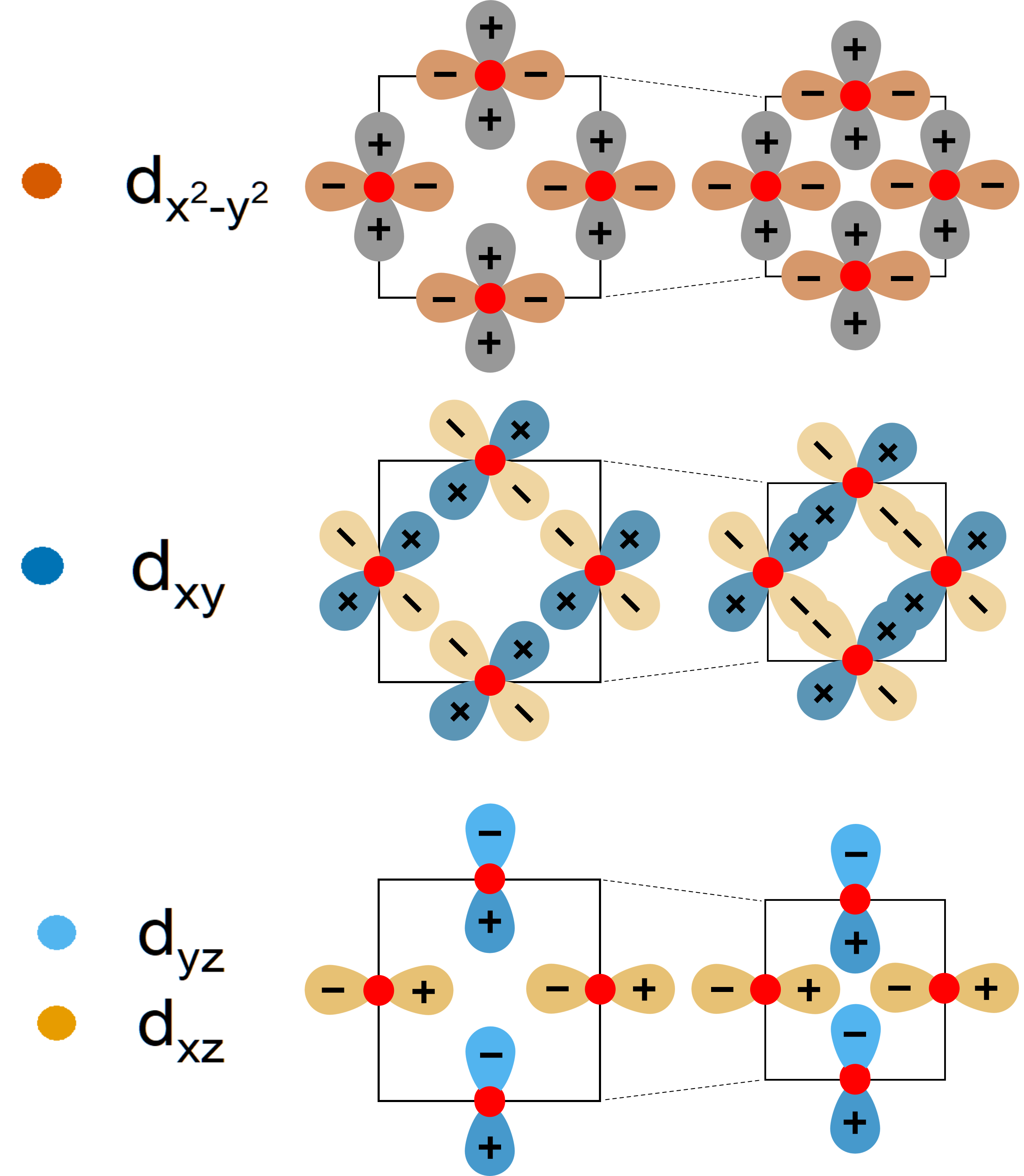}
\caption{Top-down view of d-orbitals of Mn III ions where the page plane is the epitaxial strain plane.}
\end{figure*}

\maketitle